\documentclass[english,12pt]{article}
\pdfoutput=1

\usepackage[T1]{fontenc}
\usepackage[english]{babel}
\usepackage{lmodern}
\usepackage{comment}
\usepackage{makeidx}
\usepackage{bm}
\makeindex
\usepackage[usenames,dvipsnames]{xcolor}
\usepackage[protrusion=true,expansion=true]{microtype}
\usepackage{graphicx}
\usepackage[font=small,labelfont=bf]{caption}
\usepackage{wrapfig}
\usepackage{float}
\usepackage[bottom]{footmisc}
\usepackage{booktabs}
\usepackage{multirow}
\usepackage{subfig}
\usepackage{geometry}
\usepackage{lipsum}
\usepackage[24hr]{datetime}
\usepackage{titlesec,titletoc}
\usepackage{hyperref}
\definecolor{gray75}{gray}{0.75}
\usepackage{amsmath}
\usepackage{amsfonts}
\usepackage{graphicx}
\usepackage[font=small,labelfont=bf]{caption}
\usepackage{wrapfig}
\usepackage{float}
\usepackage[bottom]{footmisc}
\usepackage{booktabs}
\usepackage{multirow}
\usepackage{subfig}
\usepackage[makeroom]{cancel}
\usepackage{geometry}
\usepackage{amsfonts}
\usepackage{hyperref}
\usepackage{color}
\usepackage{url}
\usepackage{authblk}

\makeatletter
\g@addto@macro{\UrlBreaks}{\UrlOrds}
\makeatother

\title{Reducing a cortical network to a Potts model yields storage capacity estimates}
\author[1,2,*,\P]{Michelangelo Naim}
\author[1,*]{Vezha Boboeva}
\author[1,3,4]{Chol Jun Kang}
\author[1,5]{Alessandro Treves}
\affil[1]{\footnotesize SISSA - International School for Advanced Studies, Via Bonomea 265, 34136 Trieste, Italy}
\affil[2]{\footnotesize La Sapienza Universit\`a di Roma, Piazzale Aldo Moro, 5, 00185 Roma, Italy}
\affil[3]{The Abdus Salam International Centre for Theoretical Physics, Strada Costiera 11, 34151 Trieste, Italy}
\affil[4]{\footnotesize Current address: Institute for Theoretical Physics, Department of Physics, Kim Il Sung University, Pyongyang, DPRK}
\affil[5]{\footnotesize Kavli Institute for Systems Neuroscience/Centre for Neural Computation, Norwegian University of Science and Technology, Trondheim, Norway}

\dottedcontents{section}[1.5em]{\bfseries}{1.3em}{.6em}
\date{}

\begin{document}
\maketitle

\noindent \textit{Email}: ale@sissa.it

\noindent *: These authors contributed equally to the work

\noindent \P: Current address: Weizmann Institute, Rehovot, Israel

\newpage

\begin{abstract}
An autoassociative network of Potts units, coupled via tensor connections, has been proposed and analysed as an effective model of an extensive cortical network with distinct short- and long-range synaptic connections, but it has not been clarified in what sense it can be regarded as an effective model. We draw here the correspondence between the two, which indicates the need to introduce a local feedback term in the reduced model, i.e., in the Potts network. An effective model allows the study of phase transitions. As an example, we study the storage capacity of the Potts network with this additional term, the local feedback
$w$, which contributes to drive the activity of the network towards one of the stored patterns. The storage capacity calculation, performed using replica tools, is limited to fully connected networks, for which a Hamiltonian can be defined.
To extend the results to the case of intermediate partial connectivity, we also derive the self-consistent signal-to-noise analysis for the Potts network; and finally we discuss implications for semantic memory in humans.
\end{abstract}

\newpage
\tableofcontents

\newpage
\section{Introduction}

Considerable research efforts in recent years have been driven by the ambition to reconstruct and simulate in microscopic detail the structure of the human brain, possibly at the 1:1 scale, with outcomes that have been questioned \cite{schneider2017}. A complementary perspective is that put forward by the late neuroanatomist Valentino von Braitenberg, who in many publications argued for the need to understand overarching principles of mammalian brain organisation, even by recourse to dramatic simplification \cite{braitenberg2013anatomy}. In this spirit, over 40 years ago Braitenberg proposed the notion of the \emph{skeleton} cortex, that is comprised solely of its $\mathcal{N}$ pyramidal cells \cite{braitenberg1974thoughts}. On their apical dendrites they receive predominantly synapses from axons that originate in the pyramidal cells of other cortical areas and travel through the white matter, while on their basal dendrites they receive mainly synapses from local axon
collaterals, and the two systems, A(pical) and B(asal), can be estimated to include similar numbers of synapses $C_A$ and $C_B$ per receiving cell. Braitenberg then detailed what could have later been called a \emph{small world} scheme \cite{braitenberg1978cortical}. In such a scheme, the $\mathcal{N}$ pyramidal cells are allocated to $N=\sqrt{\mathcal{N}}$ modules, each including $N$ cells, fully connected with each other -- so that $C_B=N-1$. Each cell would further receive, on the A system, $N-1$ connections from one cell drawn at random in each of the other modules, so that also $C_A=N-1$. Therefore each cell gets $2(N-1)$ connections from other pyramidal cells, the A and B systems are perfectly balanced, and the average minimal path length between any cell pair is just below 2. Of course, the modules are largely a fictional construct, apart from special cases, or at least their generality and character are quite controversial \cite{mountcastle1997columnar}, \cite{rakic2008confusing}, \cite{kaas2012evolution}, but the distinction between long-range and local connections is real, and the simple model recapitulates a rough square-root scaling of both systems, with $N\sim 10^3 \div 10^5$, in skeleton cortices which in mammals range from ca. $\mathcal{N}\sim 10^6$ to  ca. $\mathcal{N}\sim 10^{10}$.

The functional counterpart to the neuroanatomical scheme is the notion of Hebbian associative plasticity \cite{hebb2005organization}, considered as the key mechanism that modulates both long- and short-range connections between pyramidal cells. In such a view, autoassociative memory storage and retrieval are universal processes through which both local and global networks operate \cite{braitenberg2013anatomy}. Cortical areas across species would then share these universal processes, whereas the information they express would be specific to the constellation of inputs each area receives, which the simplified skeleton model does not attempt to describe. Underlying the diversity of higher-order processes of which cortical cognition is comprised, there would be the common associative operation of multi-modular autoassociative memory.

At a more abstract mathematical level, the Hopfield model of a simple autoassociative memory network \cite{hopfield1982neural} has opened the path to a quantitative statistical understanding of how memory can be implemented in a network of model neurons, through thorough analyses of attractor neural networks. Crucially, it has allowed to sketch a phase diagram, and to approach the nature of the \emph{phase transitions} an associative memory network may demonstrate, what is beyond the reach of non quantitative models. The initial analyses, with networks of binary units, then shifted towards networks with more of the properties seen in the cortex \cite{treves1990graded}, \cite{amit1995learning}.

As for modelling cortical connectivity, attempts to reproduce quantitative observations \cite{hellwig2000quantitative}, given the apparent lack of specificity at the single cell level \cite{kalisman2005neocortical}, in some cases have led to models in which, even without modules, the probability of pyramidal-to-pyramidal connections depends on the distance between neurons, rapidly decreasing beyond a distance that conceptually corresponds to the radius of a module \cite{roudi2004associative}. In other models, the basis is a strict parcellation into modules, but either with the specific assumptions of binary synapses \cite{dubreuil2016storing}
or with the network interactions across modules different in nature from those within a module (which itself can be structured in sub-modules, or \emph{mini-columns}, with winner-take-all competition among them, and synergy among fully equivalent "clone" units within them \cite{Lansner2007,Lansner2010}), thereby departing from the Braitenberg
assumption that associative Hebbian plasticity governs both intra- and inter-module interactions.

But has Braitenberg's suggested simplification, the skeleton of units with their A and B system, both associative, enabled the use of the powerful statistical-physics-derived analyses that had been successfully applied to the Hopfield model? Has it allowed an understanding of phase transitions? Only up to a point. Studies of multi-modular network models including full connectivity within individual modules and sparse connectivity with other modules could only be approached in their most basic formulation, in which all modules participate in every memory, and their sparse connectivity is random \cite{o1992short}, \cite{o1992simplest}; and attempts to articulate them further have led to analytical complexity \cite{laurogrotto1994} \cite{mari1998modeling}, \cite{levy1999associative}, \cite{mari2004extremely} or to the recourse to very sparse, effectively local coding schemes \cite{dubreuil2016storing}, without yielding a plausible quantification of storage capacity. The Potts associativ
 e network, in contrast, has been, from the early study by Ido Kanter \cite{kanter1988potts} fully analysed in its original and sparsely coded versions \cite{bolle1991stability}, \cite{bolle1992thermodynamic}, \cite{bolle1993mean}, \cite{bolle1993parallel}, \cite{kropff2005storage} and it has been argued to offer an ever further simplification of a cortical network than Braitenberg's \cite{treves2005frontal}, amenable to study also its latching dynamics \cite{russo2012cortical}. The correspondence between Braitenberg's notion and the Potts model has not, however, been discussed. We do it here, with the aim of establishing a clearer rationale for using the Potts model to study cortical processes.

\vspace{3cm}

\begin{figure}[h]
\centering
\includegraphics[scale=0.35]{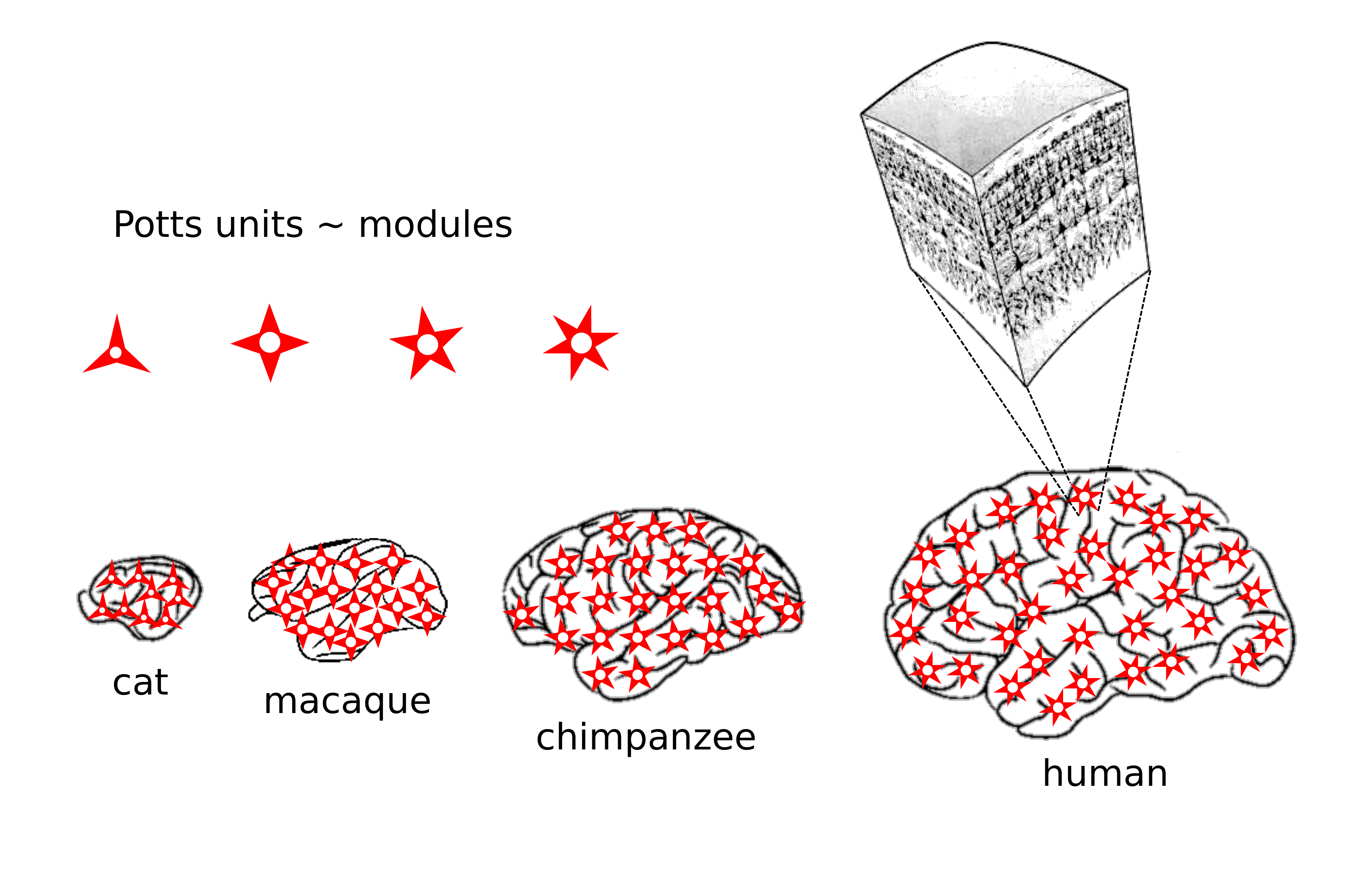}
\caption{The Braitenberg model regards a skeleton cortex of $\mathcal{N}$ pyramidal cells as comprised of $\sqrt{\mathcal{N}}$ modules of $\sqrt{\mathcal{N}}$
cells each. The Potts model then reduces each module to a multi-state unit, where a state corresponds to a dynamical attractor of the local cortical module. How should the number of states per module, $S$, be thought to scale with $\mathcal{N}$ ?}
\label{fig:Potts}
\end{figure}

\newpage
\section{The Potts network}

The Potts neural network, studied by \cite{kanter1988potts}  \cite{bolle1991stability}, \cite{bolle1992thermodynamic} and \cite{bolle1993mean}, is a network of units which can be in more than two states, generalizing Hopfield's binary network, \cite{hopfield1982neural}, in which units are either active or quiescent. A Potts unit, introduced in statistical physics in 1952 \cite{Potts1952} can be regarded in our neuroscience context as representing a local subnetwork or cortical patch of real neurons, endowed with its set of dynamical attractors, which span different directions in activity space, and are therefore converted to the states of the Potts unit (which is defined precisely as having states pointing each along a different dimension of a simplex), as schematically illustrated in Fig.~\ref{fig:Potts}. Whatever the interpretation, however, one can define the model as an
autoassociative network of $N_{m}$ Potts units interacting through tensor connections. The memories are stored in the weight matrix of the network and they are fixed, reflecting an earlier learning phase \cite{hopfield1982neural}: each memory $\mu$ is a vector or list of the states taken in the overall activity configuration by each unit $i$: $\xi^{\mu}_{i}$. We take each Potts unit to have $S$ possible active states, labelled e.g. by the index $k$, as well as one quiescent state, $k=0$, when the unit does not participate in the activity configuration of the memory. Therefore $k=0,...,S$, and each $\xi^{\mu}_{i}$ can take values in the same categorical set. The tensor weights read \cite{kanter1988potts}
\begin{equation}
\label{eqn:J_ij}
		c_{ij} J_{ij}^{kl} = \frac{c_{ij}}{c_ma(1-\frac{a}{S})} \sum\limits^{p}_{\mu=1}\left(\delta_{\xi_i^\mu k} - \frac{a}{S}\right)\left(\delta_{\xi_j^\mu l} - \frac{a}{S}\right)(1-\delta_{k0})(1-\delta_{l0}) \, ,
\end{equation}
\noindent where $i,j$ denote units, $k,l$ denote states, $a$ is the fraction of units active in each memory, $c_{ij}=1$ or 0 if unit $j$ gives input or not to unit $i$, $c_m$ is the number of input connections per unit, and the $\delta$'s are Kronecker symbols. The subtraction of the mean activity per state $a/S$ ensures a higher storage capacity \cite{kanter1988potts}. The units of the network are updated in the following way:
\begin{equation}
\sigma^k_i=\frac{\exp{(\beta r^k_i)}}{\sum^S_{l=1} \exp{(\beta r^l_i)}+\exp{[\beta (\theta^0_i+U_i)]}}
\label{eqn:sigma_i^k}
\end{equation}
\noindent and
\begin{equation}
\sigma^0_i=\frac{\exp{[\beta (\theta^0_i+U_i)]}}{\sum^S_{l=1} \exp{(\beta r^l_i)}+\exp{[\beta (\theta^0_i+U_i)]}} \, ,
\label{eqn:sigma_i^0}
\end{equation}
where $r^k_i$ is the variable representing the input to unit $i$ in state $k$ within a time scale $\tau_1$ and $U_i$ is effectively a threshold.  From Eqs.~\eqref{eqn:sigma_i^k} and \eqref{eqn:sigma_i^0}, we see that $\sum^S_{k=0} \sigma^k_i\equiv 1$, and note also that $\sigma^k_i$ takes \emph{continuous} values in the (0,1) range for each $k$, whereas the memories, for simplicity, are assumed discrete, implying that perfect retrieval is approached when $\sigma^k_i\simeq 1$ for $k=\xi^{\mu}_{i}$ and $\simeq 0$ otherwise.

\subsection{Potts model dynamics}

When the Potts model is studied as a model of cortical \emph{dynamics}, $U_i$ is often written as $U+\theta^0_i$, where $U$ is a common threshold acting on all units, and $\theta^0_i$ is the threshold component specific to unit $i$, but acting on \emph{all} its active states, and varying in time with time constant $\tau_3$. This threshold is intended to describe local inhibitory effects, which in the cortex are relayed by ${\rm GABA_A}$ and ${\rm GABA_B}$ receptors, with widely different time courses, from very short to very long. As discussed elsewhere \cite{Kang2017}, also the dynamical behaviour of the Potts model is much more interesting if both fast and slow inhibition is included. Here, however, we do not treat dynamics beyond this sketch, and stay with a single $\tau_3$ time constant for the sake of simplicity.

The time evolution of the network is governed by the equations
\begin{eqnarray}
	 \tau_1 \frac{dr^k_i(t)}{dt}&=&h^k_i(t)-\theta^k_i(t)-r^k_i(t) \nonumber \\
		\tau_2 \frac{d\theta^k_i(t)}{dt}&=&\sigma^k_i(t)-\theta^k_i(t) \label{eqn:timederivative} \\
		\tau_3 \frac{d\theta^0_i(t)}{dt}&=&\sum^S_{k=1}\sigma^k_i(t)-\theta^0_i(t) \nonumber
\end{eqnarray}
where the variable $\theta^k_i$ is a specific threshold for unit $i$ in state $k$, varying with time constant $\tau_2$, and intended to model adaptation, i.e. synaptic or neural fatigue specific to the neurons active in state $k$; and the field that the unit $i$ in state $k$ experiences reads
\begin{equation}
h^k_i=\sum^{N_m}_{j\ne i} \sum^S_{l=1} J^{kl}_{ij} \sigma^l_j+w \left(\sigma^k_i-\frac{1}{S} \sum^S_{l=1} \sigma^l_i  \right).
\label{eqn:locfield}
\end{equation}

Note that $w$ is another parameter, the ``local feedback term'', first introduced in \cite{russo2012cortical}, aimed at modelling the stability of local attractors in the full model. It helps the network converge towards an attractor, by giving more weight to the most active states, and thus it effectively deepens the attractors.

\newpage
\section{From a multi-modular Hopfield network to a Potts network}
\label{sec:HP}
We do not review here the Hopfield network \cite{hopfield1982neural} not its implementation with threshold-linear units \cite{treves1990graded} but briefly recapitulate, in order to draw the correspondence with the Potts network, the multi-modular version of the threshold-linear Hopfield network, as considered earlier without \cite{o1992short} and with globally sparse coding \cite{mari1998modeling}.

\begin{figure}[h]
\centering
\includegraphics[scale=0.5]{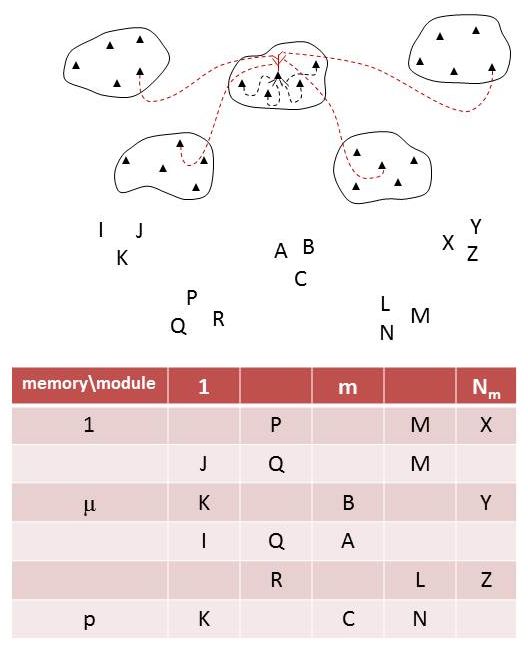}
\caption{In a cortex comprised of modules, with pyramidal cells receiving their sparse inputs from other modules on the apical dendrites (in color; top panel), memory patterns can be thought of as comprised of features, whose values are coded in the local attractors of each module (middle panel, which reproduces the layout of the modules in the top panel). Features have to be bound together by the tensor connections, in the Potts model. Sparse coding means that not all features pertain to every memory; the rest of the Potts units are in their quiescent state, as in the toy example at the bottom, where $N_m=5, p = 6, S=3, a=0.6$.}
\label{fig:ModularModel}
\end{figure}

Let us consider an underlying network of $N_{m}$ modules (\cite{o1992short}, \cite{o1992simplest}, \cite{mari1998modeling}, \cite{mari2004extremely}), each comprised of $N_{u}$ neurons, each of which is connected to all $N_{u}-1$ other neurons within the same module, and to $C_A$ other neurons distributed randomly throughout all the other modules (in earlier papers the notation $L\equiv C_A$ has been used). We make the critical ``Hopfield'' assumption \cite{hopfield1982neural} that both short- and long-range synaptic connections are symmetric. The activity $V_i$ of each neuron is a threshold-linear function of its summed input, as in \cite{treves1990graded}. The modularity finds expression in the articulation of the global activity patterns that comprise the attractor states of the network. Each module can retrieve one of $S$ local activity patterns, or \emph{features}, that are learned with the corresponding short range connections.
We index it with $\xi=1,\dots,S$. Furthermore, $p$ global activity patterns, each consisting of combinations of $aN_{m}$ features, are stored on the dilute long-range connections, as illustrated in Fig.~\ref{fig:ModularModel}. The total number of connections to a neuron is given by $C = C_A + N_{u} -1 $ and we define the fraction of long range connections as $ \gamma = C_A/C$. The model therefore partially incorporates Braitenberg's assumptions, by setting $C_B=N_u-1$; the implementation would be complete if also $N_m=N_u$, $C_A=N_m-1$ and therefore $\gamma=1/2$, but this is not necessary for the analytical treatment.

We make here the simplifying assumption that the firing rates, $\eta$, that represent a local pattern $\xi$ within a module $m$, are identically and independently distributed across units, given by the distribution $P_\eta \left( \eta_{i_m}^{\xi} \right)$. A global pattern, $\mu=1,\dots,p$, is a random combination $ \lbrace \xi_1^{\mu}, \dots , \xi_m^{\mu}, \dots, \xi_{N_{m}}^{\mu} \rbrace $, with the constraint that only $aN_m$ of the $k$'s are non-zero. We denote as $\zeta \equiv pa/S$ the average number of global patterns represented by a specific local pattern, given global sparsity $a$, and assume it for simplicity to be an integer number. We also impose, as in \cite{treves1991determines}, that $P_\eta$ satisfies $ \left\langle \eta \right\rangle = \left\langle \eta^2 \right\rangle = a_u $, such that local representations are also sparse, with sparsity parameter $a_u$ distinct from the global one $a$, both measures parametrizing, at different scales, sparse coding.

Using Hebbian covariance rules \cite{tsodyks1988enhanced} in the multi-modular network, we have
\begin{equation}\label{eqn:J_short}
	 J_{i_m,j_m}^{\rm short} = \rho_s \frac{1}{C} \sum_{\mu=1}^p  \left( \frac{\eta_{i_m}^{\xi_m^{\mu}}}{a_u} -1\right) \left(
\frac{\eta_{j_m}^{\xi_m^{\mu}}}{a_u} -1\right)
\end{equation}
\begin{equation}\label{eqn:J_long}
 J_{i_m,j_n}^{\rm long} = \rho_l \frac{c_{i_m,j_n}}{C} \sum_{\mu=1}^p  \left( \frac{\eta_{i_m}^{\mu}}{a_u} -1 \right) \left( \frac{\eta_{j_n}^{\mu}}{a_u} - 1\right)
\end{equation}
where $\rho_s$ and $\rho_l$ are parameters that adjust the dimensions of short- and long-range connections, and can regulate their relative strength. Note that we adopt the more complex index $\xi^{\mu}_m$ in Eq.~\eqref{eqn:J_short} to emphasize that the summation over $\mu$ implies repeatedly using the same local pattern $k$, for all global patterns that have $\xi^{\mu}_m=k$. The variable $c_{i_m,j_n}$ is a binary variable
\begin{equation}
	c_{i_m,j_n} =
	\left\{
	\begin{array}{lll}
	1 & \textrm{with probability} \, &\epsilon \\
       0 & \textrm{with probability} \, &(1-\epsilon)
       \end{array}
       \right.
\end{equation}
where $\epsilon = C_A/N_{u}(N_{m}-1)$.

In those cases in which an energy function can be defined, i.e., essentially, if $c_{i_m,j_n}=c_{j_n,i_m}$
the attractor states of the system, \cite{amit1992modeling}, correspond to the minima of a ``free energy''. The ``Hamiltonian'' of the multi-modular network, which is proportional to $N_u\times N_m$, is in those cases given by
\begin{eqnarray}
	\label{eqn:Hh} \mathcal{H} &=& - {1 \over 2}\sum_m \sum_{ i_m, j_m \neq i_m} J_{i_m,j_m}^{\rm short} V_{i_m} V_{j_m} -   {1 \over 2}\sum_{m,n \neq m} \sum_{ i_m, j_n} J_{i_m,j_n}^{\rm long} V_{i_m} V_{j_n} \\ \nonumber
	& =& \mathcal{H}_s + \mathcal{H}_l.
\end{eqnarray}

\subsection{Thermodynamic correspondence}

Estimating $c_{i_m,j_n}$ with its mean $\epsilon$, we can rewrite the second term above as
\begin{eqnarray*}
	 \mathcal{H}_l &=& - \sum_{m,n > m} \sum_{ i_m, j_n} J_{i_m,j_m}^{\rm long} V_{i_m} V_{j_n}  \\
	  & =& - \rho_l  \sum_{m,n > m} \sum_{ i_m, j_n} \frac{c_{i_m,j_n}}{C} \sum_{\mu=1}^p  \left( \frac{\eta_{i_m}^{\mu}}{a_u} -1 \right) \left( \frac{\eta_{j_n}^{\mu}}{a_u} - 1\right) V_{i_m} V_{j_n} \\
	  & \simeq& - \rho_l \frac{\epsilon}{C} \sum_{m,n > m} \sum_{\mu} \sum_{i_m, j_n} \left( \frac{\eta_{i_m}^{\mu}}{a_u} -1 \right) \left( \frac{\eta_{j_n}^{\mu}}{a_u} - 1\right) V_{i_m} V_{j_n}.
\end{eqnarray*}
We note that for a given pattern $\mu$ the only contribution to $\eta_{i_m}^{\mu}$ is $\eta_{i_m}^{\xi_m^{\mu}}$. We now define the local correlation of the state of the network with each local memory pattern as
\begin{equation}
\sigma_{m}^{\xi_{m}^{\mu}} = \frac{1}{N_u} \sum_{i_m} \bigg( \frac{\eta_{i_m}^{\xi_m^{\mu}}}{a_u} - 1 \bigg) V_{i_m}
\label{eqn:overlapStatic}
\end{equation}
where to avoid introducing additional dimensional parameters, we assume that the activity $V_i$ of each model neuron is measured in such units, and suitably regulated by inhibition, that the local correlations are automatically normalized to reach a maximum value of 1. We then obtain
\begin{eqnarray}
\mathcal{H}_l &=& - \rho_l \frac{\epsilon N_{u}^2}{C} \sum_{m,n > m} \sum_{\mu} \sigma_m^{\xi_m^{\mu}} \sigma_n^{\xi_n^{\mu}} \nonumber \\
 & =& - \rho_l \frac{\epsilon N_{u}^2}{C} \sum_{m,n > m} \sum_{\mu} \sum_k \sum_l \delta_{\xi_m^{\mu} k} \delta_{\xi_n^{\mu} l } \sigma_m^{k} \sigma_n^{l} \nonumber \\
 & =& - N_u\sum_{m,n > m} \sum_{k,l} J_{m n }^{k l} \sigma_m^{k} \sigma_n^{l} \, ,
\label{eqn:HL}
\end{eqnarray}
where we have introduced
\begin{equation}
J_{m n }^{k l} = \rho_l \frac{\epsilon N_{u}}{C} \sum_{\mu} \delta_{\xi_m^{\mu} k } \delta_{\xi_n^{\mu} l }= \rho_l \frac{\gamma}{N_m-1} \sum_{\mu} \delta_{\xi_m^{\mu} k } \delta_{\xi_n^{\mu} l }.
\end{equation}
On the other hand, using Eq.~\eqref{eqn:overlapStatic}, the first term can be rewritten as
\begin{eqnarray}
	 \mathcal{H}_s & =& - \sum_m \sum_{ i_m, j_m > i_m} J_{i_m,j_m}^{S} V_{i_m} V_{j_m} \nonumber\\
	  & \simeq& - \rho_s  \frac{\zeta}{C}\sum_m \sum_{ i_m, j_m > i_m} \sum_{\xi=1}^S  \left( \frac{\eta_{i_m}^{\xi}}{a_u} -1\right) \left( \frac{\eta_{j_m}^{\xi}}{a_u} -1\right) V_{i_m} V_{j_m} \nonumber\\
	    & =& - \rho_s \frac{\zeta}{C} \sum_m \sum_{\xi=1}^S \Bigg\lbrace \sum_{i_m, j_m} \left( \frac{\eta_{i_m}^{\xi}}{a_u} -1\right) \left( \frac{\eta_{j_m}^{\xi}}{a_u} -1\right) V_{i_m} V_{j_m} - \sum_{i_m} \left[ \left( \frac{\eta_{i_m}^{\xi}}{a_u} -1\right) V_{i_m} \right]^2 \Bigg\rbrace \nonumber\\
	     & \simeq& - \rho_s \frac{\zeta }{C} \sum_m \Bigg\lbrace N_{u}^2 \sum_k   \left( \sigma_m^{k} \right)^2 -
\frac{S(1-a_u)}{a_u}\sum_{i_m} \left[ V_{i_m} \right]^2 \Bigg\rbrace \, .
	\label{eqn:HS}
\end{eqnarray}
where we have noted the absence of self-interactions, and estimated with its mean $\zeta \equiv pa/S$ the number of contributions to the encoding of each local attractor state. Putting together Eqs.~\eqref{eqn:HL} and \eqref{eqn:HS}, where we neglect the last term in the $N_u\to \infty$ limit, and noting that $N_u/C \simeq C_B/C = 1-\gamma$, we have
\begin{equation}
\centering
	\mathcal{H} \simeq - N_u \sum_{m,n > m} \sum_{k,l} J_{m n }^{k l} \sigma_m^{k} \sigma_n^{l} - N_u \rho_s\zeta (1-\gamma) \sum_m \sum_k    \left( \sigma_m^{k} \right)^2 \, .
\label{eqn:Htot}
\end{equation}

We have therefore expressed the Hamiltonian of a multi-modular Hopfield network in terms of {\it mesoscopic} parameters, the $\sigma^k_m$'s, characterizing the state of each module in terms of its correlation with locally stored patterns. This could be regarded as (proportional to) the effective Hamiltonian of a reduced Potts model, if due attention is paid to entropy and temperature.

First, let us consider the temperature. Since the  $\sigma^k_m$'s are infinite (in the $N_m\to\infty$ limit) but infinitely fewer than the $V_i$'s (in the $N_u\to\infty$ limit), the correct Potts Hamiltonian is akin to a free-energy for the full multimodular model, it should scale with $N_m$ and not with $N_m\times N_u$, and it should include the proper entropy terms. One can write
\begin{equation}
\centering
\exp -\beta_{Potts}\mathcal{H}_{Potts}(\{\sigma^k_m\}) = \sum_{\{V_i\}}\exp -\beta\mathcal{H}_{}(\{V_i\}|\{\sigma^k_m\}) .
\label{eqn:free-E}
\end{equation}
The correct scaling of the Potts Hamiltonian implies that an extra $N_u$ factor present in the original Hamiltonian has to be reabsorbed in the effective inverse Potts temperature $\beta_{Potts}$, which then diverges in the thermodynamic limit. This means that the Potts network can be taken to operate at zero temperature, in relation to its interactions between modules. Within modules, however, the effects of a non-zero noise level in the underlying multi-modular network persist in the entropy terms.

Let us now turn, then, to the entropy. Here, delineating the correspondence requires suitable assumptions on the distribution of microscopic configurations that dominate the thermodynamic (mesoscopic) state of each module, which are expressed as entropy terms of the effective Potts model. One such assumption is that a module is mostly in states fragmented into competing \emph{domains} of $n_0, n_1, \dots, n_k, \dots, n_S$ units, fully correlated with the corresponding local patterns, except for the
first $n_0$, which are at a spontaneous activity level.
This would imply that, dropping the module index $m$, $\sigma^k = n_k/N_u$, and the constraint
$\sum_{k=0}^S \sigma^k=1$ is automatically satisfied. The number of microscopic states characterized by the same $S+1$-plet $n_0, \dots, n_k, \dots, n_S$  is $N_u!/ \prod_{k=0}^S n_k!$. The log of this number, which can be estimated as $-N_u \sum_{k=0}^S \sigma^k \ln \sigma^k $, has to be divided by $\beta$ and then subtracted for each module from the original Hamiltonian, as the entropy term that comes from the microscopic free-energy. This becomes the effective Hamiltonian of the Potts network by further dividing by $N_u$, because a factor $N_u$ has to be reabsorbed into $\beta$. Therefore one finds the additional entropy term in the reduced Hamiltonian
\begin{equation}
\centering
\beta \mathcal{H}_{Potts}^{\rm entropy}(\{\sigma^k_m\}) = \sum_m\sum_{k=0}^S  \sigma^k_m \ln \sigma^k_m .
\label{eqn:entropy}
\end{equation}
The above shows that the original inverse temperature $\beta$ retains its significance as a local parameter, that modulates the stiffness of each module or Potts units, even though the effective noise level in the long-range interactions between modules vanishes. The precise entropy formula depends also on the assumptions that all microscopic states be dynamically accessible from each other, which would have to be validated depending on the dynamics assumed to hold within each module. An alternative assumption is that individual units can in practice only be exchanged between a fragment correlated with local pattern $k$ and the pool $n_0$ of uncorrelated units. Under that assumption the entropy can be estimated from the log of the number $\prod_{k=1}^S (N_u!/n_0!n_k!)$, which yields
\begin{equation}
\centering
\beta \mathcal{H}_{Potts}^{\prime {\rm entropy}}(\{\sigma^k_m\}) = \sum_m\sum_{k=1}^S  \left\{\sigma^k_m \ln
\frac{\sigma^k_m }{\sigma^k_m + \sigma^0_m}+ \sigma^0_m \ln \frac{\sigma^0_m }{\sigma^k_m + \sigma^0_m}\right\}
\label{eqn:entropy_ele}
\end{equation}
as in \cite{russo2012cortical}, Eq.~(11).

Note that, in Eq.~\eqref{eqn:Htot}, the sparse connectivity between modules of the multi-modular network does not translate into a diluted Potts connectivity: each module, or Potts unit, receives inputs from each of the other $N_m-1$ modules, or Potts units. One can consider cases in which, instead, there are only $c_m$ connections per Potts unit, e.g. the \emph{highly diluted} and \emph{intermediate connectivity} considered in the storage capacity analysis below.

\subsection{Where it gets vague: inhibition and dynamics}

These arguments indicate how the local attractors of each module can be reinterpreted as dynamical variables of a system of interacting Potts units. The correspondence cannot be worked out completely, however (and Eq.~\eqref{eqn:Htot} is not fully equivalent to the Hamiltonian defined in \cite{russo2012cortical}), if anything because the effects of inhibition cannot be included, given the inherent asymmetry of the interactions, in a Hamiltonian formulation. In the body of work on neural networks stimulated by the Hopfield model, some of the effects ascribed to inhibition have been regarded as incapsulated in the peculiar \emph{Hebbian} learning rule that determines the contribution of each stored pattern to the synaptic matrix, with its subtractive terms. Similar subtractive terms can be argued on the same basis to take into account inhibitory effects at the \emph{module level}, and they lead to replace the interaction
\begin{equation}
 J_{m n }^{k l} =\rho_l \frac{\gamma}{N_m-1} \sum_{\mu} \delta_{\xi_m^{\mu} k } \delta_{\xi_n^{\mu} l }
\end{equation}
with
\begin{equation}\label{eq:tensor_subtract_adivS}
 J_{m n }^{\prime k l} =\rho_l \frac{\gamma}{N_m-1} \sum_{\mu} (\delta_{\xi_m^{\mu} k }-a/S) (\delta_{\xi_n^{\mu} l }-a/S),
\end{equation}
the form which appears in \cite{russo2012cortical}. The  local feedback term there, parametrized by $w$, can be made to roughly correspond to the second term in Eq.~\eqref{eqn:Htot} by imposing that $\rho_s \zeta (1-\gamma) / \rho_l \gamma = w/2 $.

To extend further the approximate correspondence, beyond thermodynamics and into dynamics, we may assume that
underlying the Potts network there is in fact a network of $N_m \times N_{u}$ integrate-and-fire model neurons, emulating the dynamical behaviour of pyramidal cells in the cortex, as considered by \cite{treves1993mean} and \cite{battaglia1998stable}.
The simple assumptions concerning the connectivity and the synaptic efficacies are reflected in the fact that the inputs to any model neuron in the extended network are determined by globally defined quantities, namely the mean fields, which are weighted
averages of quantities that measure, as a function of time, the effective fraction of synaptic conductances ($g$, in suitable units normalized to $\Delta g$) open on the membrane of any cell of a given class, or cluster (G) by the action of all presynaptic cells of another given class, or cluster (F)
\begin{equation}
	z_{G}^F \left(t\right) = \frac{1}{N_{local,F}} \sum_{\alpha \in F} \frac{g^{\alpha} \left(t\right)}{\Delta g_{G}^F} \, ,
\end{equation}
where $g^{\alpha}$ is the conductance of a specific synaptic input. The point is that among the clusters that have to be defined in the framework of Ref.\cite{treves1993mean}, many cluster pairs (F,G), those that comprise pyramidal cells, share the same or a similar biophysical time constant, describing their conductance dynamics  \cite{treves1993mean}, i.e.
\begin{equation}
 \frac{d z_{G}^F \left(t\right)}{dt} = - \frac{1}{\tau_G^F} z_{G}^F \left(t\right) + \nu_F \left(t -\Delta t \right) \, ,
\end{equation}
where $\nu_F \left(t\right)$ is the firing rate. If $\tau_G^F$ is the same across distinct values for $F$ and $G$, one can compare the equation for any such cluster pair to the first equation of Eq.~\eqref{eqn:timederivative}, namely
\begin{equation*}
	\tau_1 \frac{dr^k_i(t)}{dt}=h^k_i(t)-\theta^k_i(t)-r^k_i(t) \, .
\end{equation*}
Since $r^k_i$ is the temporally integrated variable representing the activity of unit $i$ in state $k$ varying with the time scale of $\tau_1$, it can be taken to correspond to the (integrated) activation of pyramidal cells in a module. One can conclude that $\tau_1$ summarizes the time course of the conductances opened on pyramidal cells by the inputs from other pyramidal cells. It represents the inactivation of synaptic conductance and, like the firing rates are a function of the $z$, our overlap is a function of the $r$. Neglecting adaptation ($\theta^k_i$), we can think of the correspondence as
\begin{equation}
	h^k_i \sim \sum_{\alpha \in F} \nu^{\alpha} \rightarrow r^k_i \sim \sum_{\alpha \in F} z^{\alpha}
\end{equation}
therefore $r^k_i$ represents the state of the inputs to the integrate-and-fire neurons within a module, i.e., a Potts unit, and we can identify the constant $\tau_1$ with the inactivation time constant for the synapses between pyramidal cells, $\tau_E^E$, whereas inhibitory and adaptation effects will be represented by $\tau_2$ and $\tau_3$ in the Potts model.

The considerations in this subsection appear to be particularly \emph{ad hoc}, as they are bound to be, since we are drawing a possible correspondence which is not a one to one mapping, but rather a reduction to a system with a very large number of variables from another system with a yet much larger number, which itself is intended to represent in simplified form the extreme complexity of the cortex. Still, the correspondence, even though approximate, helps in interpreting the result of the mathematical analysis of the "thermodynamics" of the reduced model, to which we turn next.

\newpage
\section{Storage capacity of the Potts network}

In the previous section, we have expressed the approximate equivalence between the Hamiltonian of a multi-modular Hopfield network and that of the Potts network. This means that we can study the retrieval properties of the Potts network, as an effective model of the full multi-modular network.

\subsection{Fully connected network}

In this subsection, we study the storage capacity of the Potts network with full connectivity using the classic replica method. We quantify the storage load with $\alpha\equiv p/c_m$ or, in the case of full connectivity, $\alpha\equiv p/N$. Taking inspiration from \cite{kropff2005storage} and \cite{russo2012cortical}, let us consider the Hamiltonian which is defined as:
\begin{equation}
	\mathcal{H} = -\frac{1}{2}\sum_{i,j \neq i}^N \sum_{k,l = 0}^S J_{ij}^{kl} \delta_{\sigma_i k} \delta_{\sigma_j l} + U \sum_i^N \left( 1-\delta_{\sigma_i 0} \right) - \frac{w}{2} \sum_i^N \left[ \: \sum_{k>0} \delta_{\sigma_i k}^2 - \frac{1}{S} \left(1-\delta_{\sigma_i 0}\right)^2 \right] \, .
	\label{eqn:mainham}
\end{equation}

The coupling between the state $k$ in unit $i$ and the state $l$ in unit $j$ is a Hebbian rule (\cite{kropff2005storage}, \cite{hopfield1982neural}, \cite{bolle1993mean}, \cite{russo2012cortical}, \cite{kropff2007complexity})
\begin{equation}
	\centering
	\left\{
	\begin{array}{c}
		J_{ij}^{kl} = \frac{1}{Na \left( 1 - \tilde{a} \right)} \sum_{\mu=1}^p v_{\xi^\mu_ik} v_{\xi^\nu_jl} \\
		\\
		v_{\xi^\mu_ik}=\left( \delta_{\xi_{i}^\mu k} - \tilde{a} \right)  \left( 1 - \delta_{k 0} \right)
		\end{array}
		\right.
\label{eqn:mainJ}
\end{equation}
where $N$ is the total number of units in our Potts network (for clarity we drop henceforth the subscript $N_m$, except when discussing parameters in Sect.\ref{param}),  $p$ is the number of stored random patterns, $a$ is their sparsity, i.e., the fraction of active Potts units in each, and $\tilde{a} = a/S$.
As mentioned above, $U$ is the time-independent threshold acting on all units in the network, as in \cite{kropff2005storage}. The main difference with the analysis in \cite{kropff2005storage} is that here we have included the term proportional to $w$ in Eq.~\eqref{eqn:mainham}. This self-reinforcement term pushes each unit into the more active of its states, thus providing positive feedback.

The patterns to be learned are drawn from the following probability distribution (\cite{kropff2005storage}, \cite{russo2012cortical}, \cite{kropff2007complexity})
\begin{eqnarray}
\left\{
          \begin{array}{l}
	 P \left( \xi_i^\mu = 0 \right) = 1 - a  \\
	 P_k \equiv P \left( \xi_i^\mu = k \right) = \tilde{a} \equiv a/S \, .
	 \end{array}
	 \right.
	 \label{eqn:Pk}
\end{eqnarray}
Using the trivial property that $\delta^2_{i,j}=\delta_{i,j}$ we can rewrite the Hamiltonian as
\begin{eqnarray*}
 		\mathcal{H}& = &  -\frac{1}{2Na \left( 1 - \tilde{a} \right)}\sum_{\mu=1}^p \left( \sum_{i}^N v_{\xi_{i}^\mu \sigma_i} \right)^2 + \frac{1}{2Na \left( 1 - \tilde{a} \right)}\sum_{i}^N \sum_{\mu=1}^p v_{\xi_{i}^\mu \sigma_i}^2 + \nonumber \\
 		& +& \left( U - \frac{w\left( S-1 \right)}{2 S} \right) \sum_i^N \frac{v_{\xi_{i}^\mu \sigma_i}}{\delta_{\xi_{i}^\mu \sigma_i}- \tilde{a}} 	\, .
\end{eqnarray*}
In the following let us define
\begin{equation}
\tilde{U} = U - \frac{w\left( S-1 \right)}{2 S}
\label{eqn:Utilde} \, .
\end{equation}
We now apply the replica technique (\cite{sherrington1975solvable, mezard1987spin, geszti1990physical}) to compute from $\mathcal{H}$ a free energy expressed in terms of \emph{overlap} order parameters $m^{\mu}$, following refs. \cite{amit1995learning, kanter1988potts, amit1992modeling, amit1985spin, amit1987statistical}. The $m$'s measure the correlation between the thermodynamic state of the network and each of the stored memory patterns, and we are interested here in the case where only one such order parameter (pertaining to the so-called \emph{condensed} pattern) differs from zero. The free energy of $N$ Potts units in replica theory reads
\begin{equation}
	\centering
	f = - \frac{1}{\beta} \lim_{n \rightarrow 0} \lim_{N \rightarrow \infty} \frac{\Big\langle Z^n \Big\rangle - 1}{Nn} \, ,
	\label{eqn:ftrick}
\end{equation}
where $\langle \cdot \rangle$ is an average over the quenched disorder (in this case represented by all the other, \emph{uncondensed} patterns in our network), as in \cite{amit1992modeling}. The quenched average requires introducing additional conjugate order parameters $q, r$, again as in \cite{amit1992modeling}, and their diagonal values $\tilde{q}, \tilde{r}$. In Appendix A we compute the replica symmetric free energy to be
\begin{eqnarray}
	f& = & \frac{a \left( 1 - \tilde{a} \right) }{2 } m^2 + \frac{\alpha}{2 \beta} \left[ \ln \left( a \left( 1- \tilde{a} \right) \right) + \ln \left( 1- \tilde{a} C \right) -\frac{\beta \tilde{a} q}{\left( 1- \tilde{a} C \right)} \right] + \nonumber \\
	 & +& \frac{\alpha \beta \tilde{a}^2}{2} \left( \tilde{r}\tilde{q}  - rq \right)  + \tilde{a}\tilde{q} \left[ \frac{\alpha}{2} + S \tilde{U} \right] + \nonumber \\
	& -& \frac{1}{\beta} \Bigg\langle \int D{\bf z} \ln \left( 1 + \sum_{\l \neq 0} \exp \left[ \beta \mathcal{H}_l^\xi \right] \right)\Bigg\rangle
	\label{eqn:freplicasymmetric}
\end{eqnarray}
where
\begin{equation}
\centering
\int Dz=\int dz \frac{\exp{(-z^2/2)}}{\sqrt{2\pi}} \, ,
\end{equation}
$C = \beta \left( \tilde{q} -q \right)$ (note that for consistency with the notation in earlier studies we use the same symbol $C$ to denote the --unrelated-- total number of connections per unit in the underlying multi-modular model) and
\begin{equation}
	\centering
	\mathcal{H}_l^\xi = m v_{\xi l} - \frac{\alpha a \beta \left( r - \tilde{r} \right)}{ 2 S^2} \left( 1 - \delta_{l 0} \right) + \sum_{k=1}^S \sqrt{\frac{\alpha r P_k}{S \left(1 - \tilde{a}\right)}} z_k v_{k l} \, .
	\label{eqn:Hxi}
\end{equation}
$C$ and $\mathcal{H}_l^\xi$ are both quantities that are typical of a replica analysis. $\mathcal{H}_l^\xi$ is the mean field with which the network affects state $l$ in a given unit if it is in the same state as condensed pattern $\xi$ (note that $\mathcal{H}_0^\xi = 0$). No such interpretation can be given to $C$: it measures the difference between $\tilde{q}$, the mean square activity in a given replica, and $q$, the coactivation between two different replicas. Note that in the zero temperature limit ($\beta \rightarrow \infty$), this difference goes to $0$, such that $C$ is always of order $1$. It will be clarified in section 4.3, through a separate analysis, that $C$ is related to the derivative of the output of an average neuron with respect to variations in its mean field.

The self-consistent mean field equations in the limit of $\beta \rightarrow \infty $ are obtained by taking the derivatives of $f$ with respect to the three replica symmetric variational parameters, $m , q , r$
\begin{align}\label{eqn:mMF}
	m & =  \frac{1}{a \left( 1 - \tilde{a} \right)} \Bigg\langle \int Dz \sum_{l \neq 0} v_{\xi l}  \left[ \frac{1}{1 + \sum\limits_{n \neq l} \exp \left[ \beta \left( \mathcal{H}_l^\xi - \mathcal{H}_n^\xi \right) \right]} \right] \Bigg\rangle \nonumber \\
	& \rightarrow  \frac{1}{a \left( 1 - \tilde{a} \right)}  \sum_{l \neq 0} \Bigg\langle \int Dz \: v_{\xi l}  \prod_{n \neq l} \Theta \left[ \mathcal{H}_l^\xi - \mathcal{H}_n^\xi \right] \Bigg\rangle
\end{align}

\begin{equation}
\centering
	q \rightarrow \tilde{q} = \frac{1}{a} \sum_{l \neq 0} \Bigg\langle \int Dz  \prod_{n \neq l} \Theta \left[ \mathcal{H}_l^\xi - \mathcal{H}_n^\xi \right] \Bigg\rangle
\label{eqn:qMF}
\end{equation}

\begin{equation}
\centering
	C = \frac{1}{\tilde{a}^2 \sqrt{\alpha r}} \sum_{l\neq0} \sum_{k} \Bigg\langle \int Dz \sqrt{\frac{P_k}{S\left( 1 - \tilde{a} \right)}} v_{k l} z_k \prod_{n \neq l} \Theta \left[ \mathcal{H}_l^\xi - \mathcal{H}_n^\xi \right] \Bigg\rangle
\label{eqn:CMF}
\end{equation}

\begin{equation}
\centering
	\tilde{r} \rightarrow r = \frac{q}{ \left( 1 - \tilde{a}C \right) ^2}
\label{eqn:rMF}
\end{equation}

\begin{equation}
\centering
	\beta \left( r - \tilde{r} \right) = 2 \bigg( \tilde{U} \frac{S^2}{a \alpha} - \frac{C}{1-\tilde{a} C} \bigg) \, .
\label{eqn:deltarMF}
\end{equation}
The $\Theta$ function gives non-vanishing contribution only for $\mathcal{H}_l^\xi - \mathcal{H}_n^\xi > 0$, i.e.
\begin{equation*}
	\sum_{k>0} \left( v_{k l} - v_{k n} \right) z_k > - m \sqrt{\frac{S^2 \left( 1 - \tilde{a} \right)}{\alpha a r}}  \left( v_{\xi l} - v_{\xi n} \right) - \frac{\alpha a \beta \left( r - \tilde{r} \right)}{2 S^2} \sqrt{\frac{S^2 \left( 1 - \tilde{a} \right)}{\alpha a r}} \left( \delta_{n 0} -\delta_{l 0} \right) \, .
\end{equation*}
Moreover, it is convenient to introduce two combinations of order parameters,
\begin{eqnarray*}
	 x& =& \frac{\alpha a \beta \left( r - \tilde{r} \right)}{2 S^2} \sqrt{\frac{S^2 \left( 1 - \tilde{a} \right)}{\alpha a r}} \, , \\
	  y& =& m \sqrt{\frac{S^2 \left( 1 - \tilde{a} \right)}{\alpha a r}} \ .
\end{eqnarray*}
At the saddle point, they become
\begin{eqnarray}
	 x &=& \frac{1}{\sqrt{q} + \tilde{a}C \sqrt{r}} \sqrt{\frac{1 - \tilde{a} }{\tilde{\alpha}}} \left[ \tilde{U} -\tilde{\alpha} \frac{C}{2} \sqrt{\frac{r}{q}} \: \right] \, , \nonumber \\
	 y& =& \sqrt{\frac{1 - \tilde{a} }{\tilde{\alpha}}} \left( \frac{m}{\sqrt{q} + \tilde{a} C \sqrt{r}} \right) \  ,
	\label{eqn:full_xy}
\end{eqnarray}
where  $ \tilde{\alpha} = \alpha a/S^2$. By computing the averages in Eqs.~\eqref{eqn:mMF} and \eqref{eqn:deltarMF}, we get three equations that close the self consistent loop with Eq.~\eqref{eqn:full_xy},
\begin{eqnarray}
	q &= & \frac{1-a}{\tilde{a}} \int D p \int_{y \tilde{a} + x - i \sqrt{\tilde{a}} p}^\infty D z \phi \left( z \right)^{S-1} \nonumber \\ \label{eqn:full_q}
	& +& \int D p \int_{-y \left( 1 - \tilde{a} \right) + x - i \sqrt{\tilde{a}} p}^\infty D z \phi \left( z + y\right)^{S-1} \\ \nonumber
	& + &\left( S-1 \right) \int D p \int_{y \tilde{a} + x - i \sqrt{\tilde{a}} p}^\infty D z \phi \left( z -y \right) \phi \left( z \right)^{S-2} \, ,
\end{eqnarray}

\begin{eqnarray}
	 m &= &\frac{1}{1-\tilde{a}} \int D p \int_{-y \left( 1 - \tilde{a} \right) + x - i \sqrt{\tilde{a}} p}^\infty D z \phi \left( z + y\right)^{S-1} - q \frac{\tilde{a}}{1-\tilde{a}} \, ,
	\label{eqn:full_m}
\end{eqnarray}

\begin{eqnarray}
	C\sqrt{r}& = & \frac{1}{\sqrt{\tilde{\alpha} \left( 1 - \tilde{a}\right)}} \Bigg\lbrace \frac{1-a}{\tilde{a}}\int D p \int_{y \tilde{a} + x - i \sqrt{\tilde{a}} p}^\infty D z \left( z + i \sqrt{\tilde{a}} p \right) \phi \left( z \right)^{S-1} \nonumber \\ \label{eqn:full_R}
	& +& \int D p \int_{-y \left( 1 - \tilde{a} \right) + x - i \sqrt{\tilde{a}} p}^\infty D z \left( z + i \sqrt{\tilde{a}} p \right) \phi \left( z + y\right)^{S-1} \\ \nonumber
	& + &\left( S-1 \right) \int D p \int_{y \tilde{a} + x - i \sqrt{\tilde{a}} p}^\infty D z \left( z + i \sqrt{\tilde{a}} p \right) \phi \left( z -y \right) \phi \left( z \right)^{S-2} \Bigg\rbrace \, ,
\end{eqnarray}
where $\phi(z)=\big(1+\textrm{erf}(z/\sqrt{2})\big)/2$. Eqs.~\eqref{eqn:full_xy}-\eqref{eqn:full_R} are complicated in their current form, such that it is useful to see their behavior in some limit cases. One such limit case is $\tilde{a}\ll 1$. Using the equalities
\begin{eqnarray*}
\int D w&=&\int \frac{dw}{\sqrt{2 \pi}} \exp{(-w^2/2)}=1\\
  d\phi&=&Dz \\
 1-\phi(x)&=&\phi(-x)
\end{eqnarray*}
and considering that $\phi(x)\sim \Theta(x)$ (the Heaviside function) away from $x\sim 0$, we get to the following self-consistent equations
\begin{eqnarray}
x&=& \frac{1}{\sqrt{\tilde{\alpha}q}}  \left( \tilde{U}-\frac{\tilde{\alpha}C}{2} \sqrt{\frac{r}{2}}  \right) \\
y&=& \frac{m}{\sqrt{\tilde{\alpha}q}}\\
m &=&\phi(y-x)\\
q&=& \frac{1-a}{\tilde{a}}\phi(-x)+\phi(y-x)\\
C\sqrt{r}&=& \frac{1}{2\pi \tilde{a}} \left\{\frac{1-a}{\tilde{a}} \exp{(-x^2/2)}+\exp{(-(y-x)^2/2)}  \right\}.
\end{eqnarray}

\subsection{Diluted networks and the highly diluted limit}

A more biologically plausible case is that of {\it diluted} networks, where the number of connections per unit $c_m$ is less than $N$.  Specifically, we consider connections of the form $c_{ij} J_{ij}$, where $J_{ij}$ is the usual symmetric matrix derived from Hebbian learning. $c_{ij}$ equals $0$ or $1$ according to a given probability distribution and we note $\lambda = \langle c_{ij} \rangle /N = c_m/N$ the dilution parameter. In general, $c_{ij}$ is different from $c_{ji}$, leading to asymmetry in the connections between units.

When the connectivity is not full, the type of probability distribution assumed for the $c_{ij}$ matters. We then consider three different distributions. The first is referred to as {\it random dilution} (RD), which is
\begin{equation}
P(c_{ij}, c_{ji}) = P(c_{ij})P(c_{ji})
\end{equation}
with
\begin{equation}
P(c_{ij}) = \lambda \delta(c_{ij}-1) + (1-\lambda) \delta(c_{ij}) \, .
\end{equation}
The second is the {\it symmetric dilution} (SD), defined by
\begin{equation}
P(c_{ij}, c_{ji}) = \lambda \delta(c_{ij}-1)\delta(c_{ji}-1) + (1-\lambda) \delta(c_{ij})\delta(c_{ji}) \, .
\end{equation}
The third is what we call {\it state dependent random dilution} (SDRD) --specific to the Potts network-- in which
\begin{equation}
P(c_{ij}^{kl}) = \lambda \delta(c_{ij}^{kl}-1) + (1-\lambda) \delta(c_{ij}^{kl}) \, ;
\end{equation}
note that in this case the connectivity coefficients are state-dependent.

We have performed simulations with all three types of connectivity, but will focus the analysis onto the RD type, which is the simplest to treat analytically. The storage capacity curve for all three models, estimated from simulations, will be shown later in Fig.~\ref{fig:SD_RD_SDRD_S2_S5}. RD and SD are known in the literature as Erdos-Renyi graphs. Many properties are known about such random graph models \cite{erdos1960evolution}, \cite{engel2004large}. It is known that for $\lambda$ below a critical value, essentially all connected components of the graph are trees, while for $\lambda$ above this critical value, loops are present. In particular, a graph with $c_m < \log(N)$ will almost surely contain isolated vertices and be disconnected, while with $c_m > \log(N)$ it will almost surely be connected.
$\log(N)$ is a threshold for the connectedness of the graph, distinguishing the highly diluted limit, for which a simplified analysis of the storage capacity is possible, as in \cite{derrida1987exactly}, from the intermediate case of the next section, for which a complete analysis is necessary, following the approach by Shiino and Fukai \cite{ShiinoandFukai1993}.

With Random Dilution, the capacity cannot be analysed through the replica method, as the symmetry of the interactions is a necessary condition for the existence of an energy function, and hence for the application of the thermodynamic formalism. We therefore apply the signal to noise analysis. The local field of unit $i$ in state $k$ writes
\begin{equation}\label{eqn:field}
h^k_i=\sum_{j} \sum_{l} c_{ij} J^{kl}_{ij} \sigma_j^l-\tilde{U} \left(1-\delta_{k,0}\right)
\end{equation}
where the coupling strength between two states of two different units is defined as
\begin{equation}\label{eqn:diluted_J}
J^{kl}_{ij}=\frac{1}{c_ma (1-\tilde{a})} \sum_\mu v_{\xi^\mu_i k}v_{\xi^\mu_j l} \, .
\end{equation}
In the highly diluted limit $c_m \sim \log(N)$ at most, the assumption is that the field can be written simply as the sum of two terms, signal and noise. While the signal is what pushes the activity of the unit such that the network configuration converges to an attractor, the noise, or the crosstalk from all of the other patterns, is what deflects the network away from the cued memory pattern.
The noise term writes
\begin{equation*}
n_i^k  \propto \sum\limits_{\mu>1}^{p} \sum\limits_{j(\neq i)}^{N} \sum\limits_{l} v_{\xi^\mu_ik} v_{\xi^\mu_jl} \, \sigma_j^l \, ,
\end{equation*}
that is, the contribution to the weights $J_{ij}^{kl}$ by all non-condensed patterns. By virtue of the subtraction of the mean activity in each state $\tilde{a}$, the noise has vanishing average:
\begin{equation*}
\langle n_i^k \rangle_{P(\xi)} \propto \sum\limits_{\mu>1}^{p} \sum\limits_{j(\neq i)}^{N} \sum\limits_{l} \langle v_{\xi_i^{\mu},k}\rangle \langle v_{\xi_j^{\mu},l} \sigma_j^l \rangle = 0 \, .
\end{equation*}
Now let us examine the variance of the noise. This can be written in the following way:
\begin{equation*}
\langle (n_i^k)^2 \rangle \propto \sum_{\mu>1}^{p} \sum\limits_{j(\neq i)=1}^{N} \sum_l \sum_{\mu'>1}^{p} \sum\limits_{j'(\neq i)=1}^{N} \sum_{l'} \langle v_{\xi_i^{\mu},k}\, v_{\xi_i^{\mu'},k}\rangle \langle v_{\xi_j^{\mu},l}\,  v_{\xi_{j'}^{\mu'},l'}\, \sigma_j^l \sigma_{j'}^{l'} \rangle \, ,
\end{equation*}
where statistical independence between units has been used. For randomly correlated patterns, all terms but $\mu = \mu '$ vanish. Having identified the non-zero term, we can proceed with the capacity analysis. We can express the field using the overlap parameter, and single out, without loss of generality, the first pattern as the one to be retrieved
\begin{equation}\label{eqn:field_fm}
h_i^k = v_{\xi^1_ik} m_i^{1} + \sum_{\mu>1} v_{\xi^\mu_ik} m_i^{\mu} - \tilde{U}(1-\delta_{k0}) .
\end{equation}
where we define the local overlap $m_i$ as
\begin{equation}\label{eqn:local_overlap}
 m_i=\frac{1}{c_m a (1-\tilde{a})}\sum_j \sum_l c_{ij}v_{\xi^1_jl}\sigma_j \, .
\end{equation}
We now write
\begin{equation}\label{eqn:ansatz_noise_diluted}
\sum_{\mu>1} v_{\xi_i^{\mu},k} m_i^{\mu} \equiv \sum_{n=1}^S v_{n,k}\, \rho^n \, z_i^n
\end{equation}
where $\rho$ is a positive constant and $z_i^n$ is a standard Gaussian variable. Indeed in highly diluted networks the l.h.s., i.e. the contribution to the field from all of the non-condensed patterns $\mu>1$, is approximately a normally distributed random variable, as it is the sum of a large number of uncorrelated quantities. $\rho$ can be computed to find
\begin{equation}
\rho^n= \sqrt{\frac{\alpha P_n }{(1-\tilde{a})S} q}
\end{equation}
where we have defined

\begin{equation}
q= \bigg\langle \frac{1}{Na}\sum_j \sum_{l} (\sigma_j^l)^2 \bigg\rangle \, .
\end{equation}
The mean field then writes
\begin{equation}
h^k_i= v_{\xi^1_i k} m + \sum_{n=1}^S v_{n,k}  \sqrt{\frac{\alpha P_n}{(1-\tilde{a})S} q} z_n -\tilde{U}(1-\delta_{k0}) \, .
\end{equation}

Averaging $m_i$ and $q$ over the connectivity and the distribution of the Gaussian noise $z$, and taking the $\beta\rightarrow\infty$ we get to the mean field equations that characterize the fixed points of the dynamics, Eqs.~\eqref{eqn:mMF} and \eqref{eqn:qMF}. In the highly diluted limit however, we do not obtain the last equation of the fully connected replica analysis, Eq.~\eqref{eqn:rMF}.

The difference between fully connected and diluted cases must vanish in the $\tilde{a}\ll 1$ limit, as shown in (\cite{kropff2005storage}, \cite{derrida1987exactly}). In this limit we have $x =\tilde{U}/\sqrt{\tilde{\alpha}q}$, $y = m/\sqrt{\tilde{\alpha}q}$ while Eqs.~\ref{eqn:full_m} and \eqref{eqn:full_q} remain identical.

\subsection{Network with partial connectivity}

We now consider the more complex case of partial connectivity, \emph{i.e.} $\log(N) < c_m < N$, which can be approached with the self-consistent signal to noise analysis (SCSNA) \cite{ShiinoandFukai1993}. As in the previous section, we can express the field using the overlap parameter, and single out the contribution from the pattern to be retrieved, that we label as $\mu=1$, as in Eq.~\eqref{eqn:field}.
With high enough connectivity, however, one must revise Eq.~\eqref{eqn:ansatz_noise_diluted}: the mean field has to be computed in a more refined way, through the SCSNA method that we recapitulate here (see also \cite{roudi2004associative}, \cite{Kropff2013}).

The noise term is assumed to be a sum of two terms
\begin{equation}\label{eqn:ansatz_noise_intermediate}
\sum_{\mu>1} v_{\xi_i^{\mu},k} m_i^{\mu} = \gamma_i^k \sigma_i^k + \sum_{n=1}^S v_{n,k} \, \rho_i^n z_i^n
\end{equation}
where $z_i^n$ are standard Gaussian variables, and $\gamma_i^k$ and $\rho_i^n$ are positive constants to be determined self-consistently.
The first term, proportional to $\sigma_i^k$, represents the noise resulting from the activity of unit $i$ on itself, after having reverberated in the loops of the network;
the second term contains the noise which propagates from units other than $i$.
The activation function writes
\begin{equation}\label{eqn:activation_function}
\sigma_i^k =  \frac{e^{\beta h_i^k}}{\sum\limits_l e^{\beta h_i^l}}  \equiv F^k\Big( \big\{y_i^l+\gamma_i^l \sigma_i^l\big\}_l\Big)\, .
\end{equation}
where $y_i^l = v_{\xi_i^{1},l} m_i^{1} + \sum_n v_{n,l} \rho_i^n z_i^n - U(1-\delta_{l,0})$.
One would need to find $\sigma_i^k$ as
\begin{equation}
\sigma_i^k = G^k\Big(\big\{y_i^l\big\}_l\Big) \ ,
\end{equation}
where $G^k$ are functions solving Eq.~\eqref{eqn:activation_function} for $\sigma_i^k$.
However, Eq.~\eqref{eqn:activation_function} cannot be solved explicitly.
Instead we make the assumption that $\{\sigma_i^l\}$ enters the fields $\{h_i^l\}$ only through their mean value $\langle \sigma_i^l\rangle$, so that we write
\begin{equation}
	 G^k\Big(\big\{y_i^l\big\}_l\Big) \simeq F^k\Big( \big\{y_i^l+\gamma_i^l \langle \sigma_i^l\rangle \big\}_l\Big) \ .
\end{equation}
We report to Appendix B the details of the calculation that yield $\gamma_i^k=\gamma$ and $\rho_i^k = \rho^k$.
\begin{equation}
\gamma=\frac{\alpha}{S} \lambda\frac{\Omega/S}{1-\Omega/S}
\end{equation}
where $\alpha = p/c_m$, $\langle\cdot\rangle$ indicates the average over all patterns and where we have defined
\begin{equation}
\Omega = \bigg\langle\frac{1}{N}\sum_{j_1} \sum_{l_1}\frac{\partial G_{j_1}^{l_1}}{\partial y^{l_1}}  \bigg\rangle\, .
\end{equation}
From the variance of the noise term one reads
\begin{equation}
(\rho^n)^2  =  \frac{\alpha P_n}{S(1-\tilde{a})} q \Big\lbrace 1+ 2\lambda \Psi + \lambda \Psi^2  \Big\rbrace \, ,
\end{equation}
where we have defined
\begin{equation}\label{eqn:q}
q = \bigg\langle \frac{1}{Na} \sum_{j,l} (G_j^l)^2 \bigg\rangle
\end{equation}
and
\begin{equation}\label{eqn:psi}
\Psi = \frac{\Omega/S}{1-\Omega/S} \ .
\end{equation}
The mean field received by a unit is then
\begin{equation}
 \mathcal{H}_k^{\xi} = v_{\xi,k} m + \frac{\alpha}{S} \lambda \Psi (1- \delta_{k,0}) + \sum_{n=1}^S v_{n,k} z^n \sqrt{\frac{\alpha P_n}{S(1-\tilde{a})} q \Big\lbrace 1+ 2\lambda \Psi + \lambda \Psi^2  \Big\rbrace} - \tilde{U}(1-\delta_{k,0}) \, .
\end{equation}

Taking the average over the non-condensed patterns (the average over the Gaussian noise $z$), followed by the average over the condensed pattern $\mu=1$ (denoted by $\langle\cdot\rangle_{\xi}$), in the limit $\beta \rightarrow \infty$, we get the self-consistent equations satisfied by the order parameters
\begin{eqnarray}\label{eqn:fixed_pt_m_q}
m &=& \frac{1}{a(1-\tilde{a})} \Bigg\langle \int D^S z \sum\limits_{l(\neq 0)} v_{\xi,l} \prod_{n(\neq l)} \Theta(\mathcal{H}_l^{\xi}-\mathcal{H}_n^{\xi}) \Bigg\rangle_{\xi} \ , \\[1ex]
q &=& \frac{1}{a} \Bigg\langle \int D^S z \sum\limits_{l(\neq 0)} \prod_{n(\neq l)} \Theta(\mathcal{H}_l^{\xi}-\mathcal{H}_n^{\xi}) \Bigg\rangle_{\xi} \ ,
\end{eqnarray}
\begin{eqnarray}\label{eqn:fixed_pt_Omega}
 \Omega &= &\Bigg\langle  \int D^S z \sum\limits_{l(\neq 0)} \sum\limits_{k} z^k \frac{\partial z^k}{\partial y^l} \prod_{n(\neq l)} \Theta(\mathcal{H}_l^{\xi}-\mathcal{H}_n^{\xi}) \Bigg\rangle_{\xi}.
\end{eqnarray}
where in the last equation for $\Omega$, integration by parts has been used. Note the similarities to Eqs.~\eqref{eqn:mMF}-\eqref{eqn:CMF}, obtained through the replica method for the fully connected case. The equations just found constitute their generalization to $\lambda < 1$. In particular, in the highly diluted limit $\lambda \rightarrow 0$,  we get $\gamma \rightarrow 0$ and $(\rho^n)^2 \rightarrow \alpha P_nq/(1-\tilde{a})S$, which are the results obtained in the previous section; in the fully
connected case, $\lambda =1$, the correspondence between the $m$ and $q$ variables is obvious, while for $\Omega$ it can be shown with some algebraic manipulation. Indeed, from the following identity,
\begin{equation}
\rho^2  =  \frac{\alpha P_n}{S(1-\tilde{a})} q (1 + \Psi)^2\, ,
\end{equation}
by using the replica variable $r = q/(1-\tilde{a}C)^2$ we get
\begin{equation}
\rho^2  =  \frac{\alpha P_n}{S(1-\tilde{a})} r (1 -\tilde{a}C)^2 (1 + \Psi)^2\ .
\end{equation}
By comparing this with Eq.~\eqref{eqn:Hxi}, the mean field, we get an equivalent expression for $\Psi$,
\begin{equation}
\Psi = \frac{\tilde{a}C}{1-\tilde{a}C} \, .
\end{equation}
From the original definition of $\Psi$ in Eq.~\eqref{eqn:psi}, it follows that the order parameter $C$, obtained through the replica method, is equivalent to $\Omega$, up to a multiplicative constant:
\begin{equation}
C = \Omega/a\, .
\end{equation}
We can show that Eq.~\eqref{eqn:fixed_pt_Omega} coincides with Eq.~\eqref{eqn:CMF}.
Moreover, by comparing the SCSNA result for $\gamma$ to the replica one, we must have
\begin{equation}
\frac{\alpha}{S} \Psi - \tilde{U} = - \frac{\alpha a \beta (r-\tilde{r})}{2S^2}
\end{equation}
from which
\begin{equation}
\beta (r-\tilde{r}) = 2 \bigg(\tilde{U}\frac{S^2}{\alpha a} - \frac{C}{1-\tilde{a}C} \bigg) \ ,
\end{equation}
identical to Eq.~\eqref{eqn:deltarMF}.

\newpage
\section{Simulation results}

\begin{figure}[H]
\begin{center}
    \begin{tabular}{cc}
      \resizebox{60mm}{!}{\includegraphics{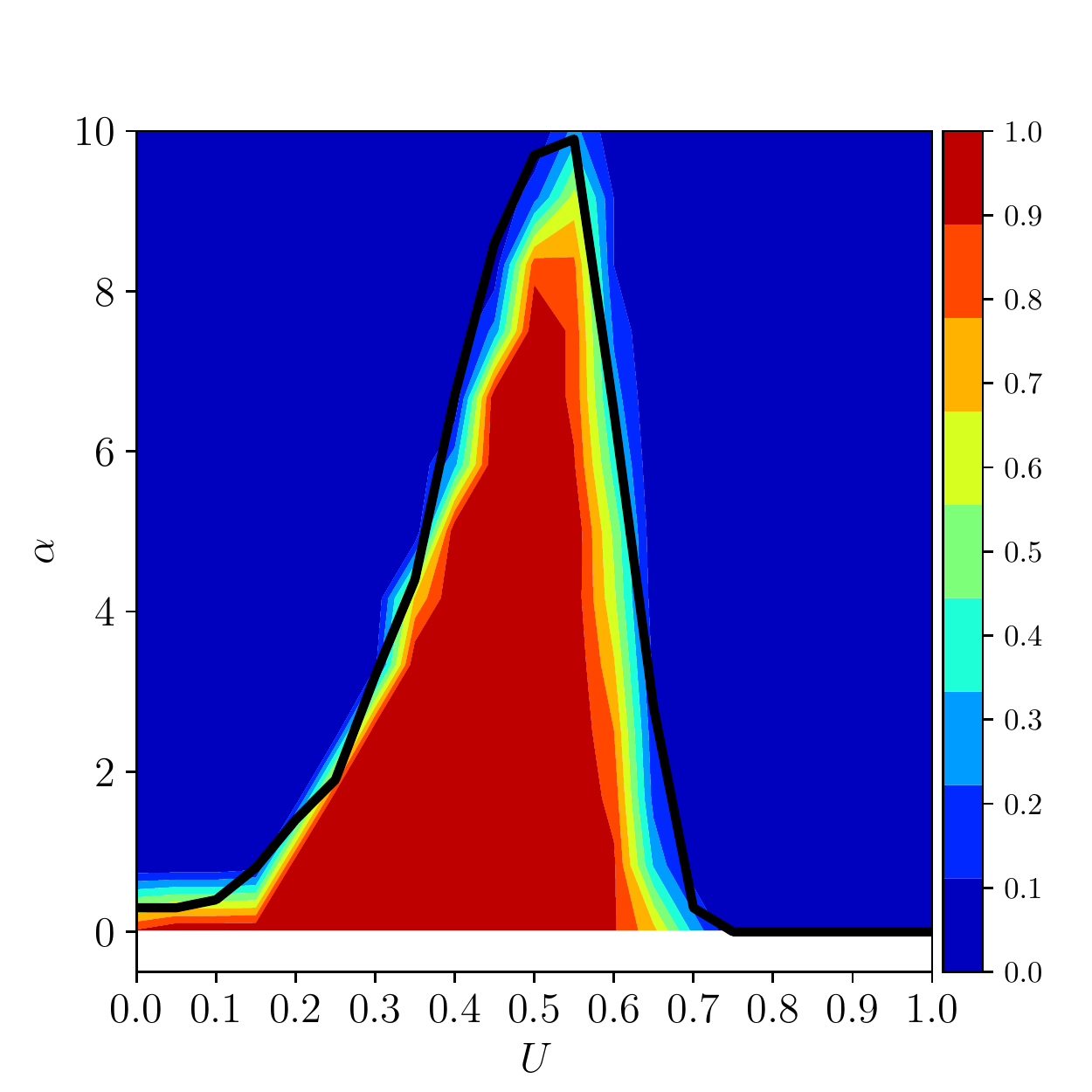}} &
      \hspace*{1cm}
      \resizebox{60mm}{!}{\includegraphics{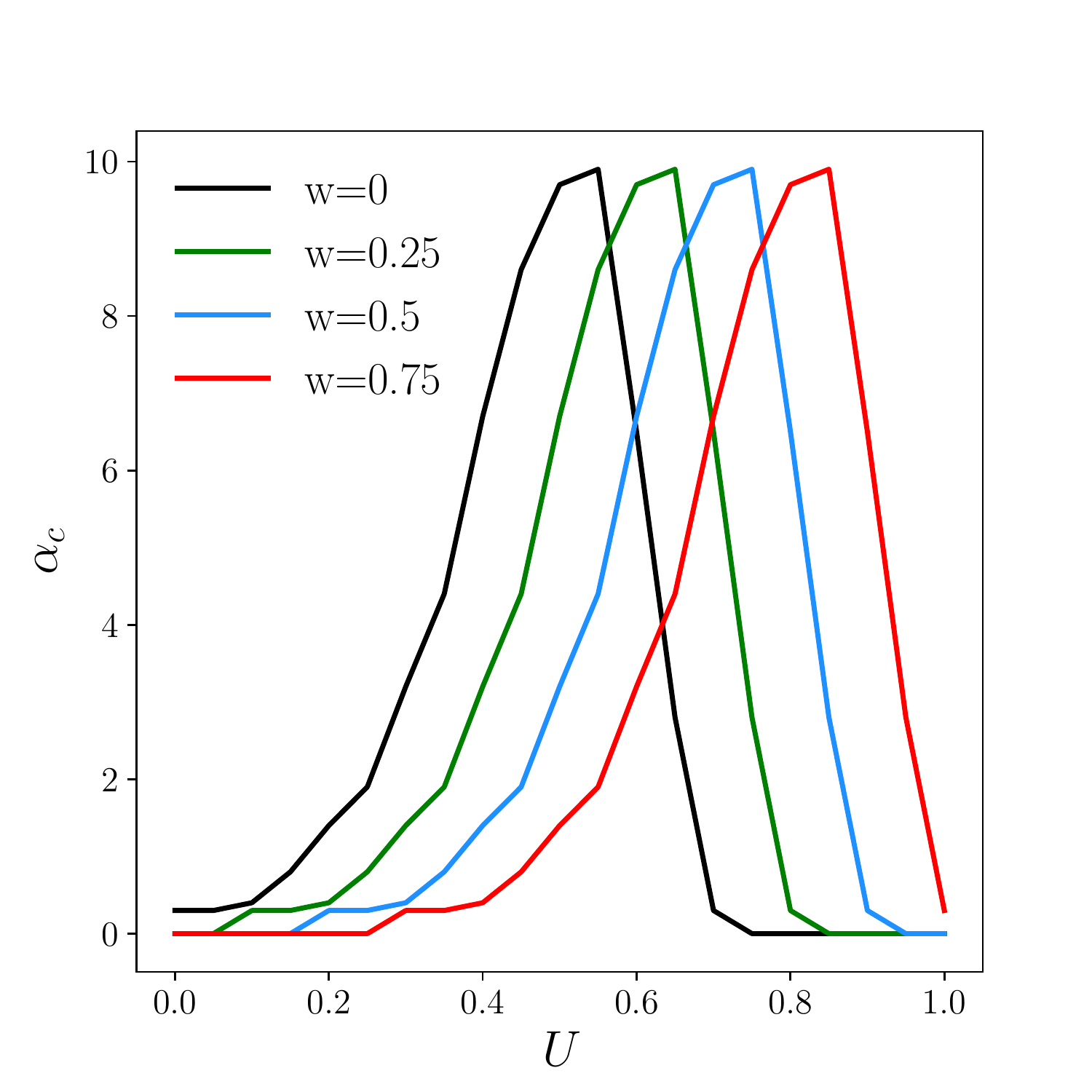}} \\
      (a) & (b)
    \end{tabular}
    \caption{(a) How often a fully connected Potts network retrieves memories, as a function of the threshold $U$ and the number of stored memories $p$, with $N=1000$, $S=7$, $a=0.25$, $\beta=200$ and $w=0.0$. Color represents the fraction of simulations in which the overlap between the activity state of network and a stored pattern is $\geq 0.9$. The solid line is obtained by numerical solution of Eqs.\eqref{eqn:full_xy}-\eqref{eqn:full_R}. (b) The dependence of $\alpha_c$ on $U$ for different values of $w$. If the threshold $U$ is already set to its optimal value, subtracting a non-zero $w$ is detrimental to the capacity, but if it can be adjusted
\emph{after} considering $w$, it can lead to an optimal effective threshold $\tilde{U}$, maximizing capacity.}\label{fig:alpha_U}
\end{center}
\end{figure}

Do computer simulations confirm the analyses above? Starting with the effect of setting the overall threshold, we show, in
Fig.~\ref{fig:alpha_U}(a), retrieval performance as a function of the threshold for $w=0.0$, both through simulations and by solving Eqs.~\eqref{eqn:full_xy}.
It is clear that the simulations agree very well with numerical results. The maximum storage capacity $\alpha_c$ (where $\alpha\equiv p/c_m$, or $\alpha\equiv p/N$ for a fully connected Potts network) is found at approximately $U=0.5$, as can also be shown through a simple signal to noise analysis. It is possible to compute approximately the standard deviation $\gamma_i^k$ of the field,
Eq.~\eqref{eqn:field} with respect to the distribution of all the patterns, as well as the connectivity $c_{ij}$, by making the assumption that all units are aligned with a specific pattern to be retrieved $\sigma_j^l = \xi^1_j$.
We further discriminate units that are in active states $\xi_i^{1}\neq0$ from those that are in the quiescent states $\xi_i^{1}=0$ in the retrieved pattern $\mu=1$.
\begin{equation}\label{eqn:sd field}
\gamma_i^k \equiv \sqrt{\langle(h_i^k)^2\rangle - \langle h_i^k\rangle^2} = \sqrt{\frac{(p-1)a}{c_m S^2} + (\delta_{\xi_i^{1}, k}-\tilde{a})^2 \Bigg(\frac{1}{c_m a} - \frac{1}{N}\Bigg)}.
\end{equation}
The optimal threshold $U_{0}$ is one that separates the two distributions, optimally, such that a minimum number of units in either distribution reach the threshold to go in the wrong state
\begin{equation*} \frac{ U_{0}- \langle h_i^k|_{\xi_i^{1}=0} \rangle}{ \gamma_i^k|_{\xi_{i}^{1}=0}} = - \frac{ U_{0}- \langle h_i^k|_{\xi_i^{1}\neq0}\rangle}{ \gamma_i^k|_{\xi_{i}^{1}\neq0} }
\end{equation*}
\begin{equation}\label{eqn:optimal threshold}
U_{0}= \frac{\gamma_i^k|_{\xi_{i}^{1}=0}}{\gamma_i^k|_{\xi_{i}^{1}=0}+\gamma_i^k|_{\xi_{i}^{1}\neq0}} - \frac{a}{S}.
\end{equation}
We can see that $U_{0} \longrightarrow 1/2 - \tilde{a}$ for  $\gamma_i^k|_{\xi_{i}^{1}=0} \sim \gamma_i^k|_{\xi_{i}^{1}\neq0}$, roughly consistent with the replica analysis and simulations in Fig.~\ref{fig:alpha_U}(a) (in fact the variance
$\gamma_i^k|_{\xi_{i}^{1}=0}$ is larger than $\gamma_i^k|_{\xi_{i}^{1}\neq0}$, especially for low $a$, hence $U_0$ is slightly
larger and has a more complex dependence on the sparsity). Given such an optimal value for $U$, Fig.~\ref{fig:alpha_U}(b) shows that the effect of the feedback term $w$ on the storage capacity, purely subtractive, is just to shift to the right the optimal value.

\subsection{The effect of network parameters}

\begin{figure}[h]
\begin{center}
    \begin{tabular}{cc}
      \resizebox{60mm}{!}{\includegraphics{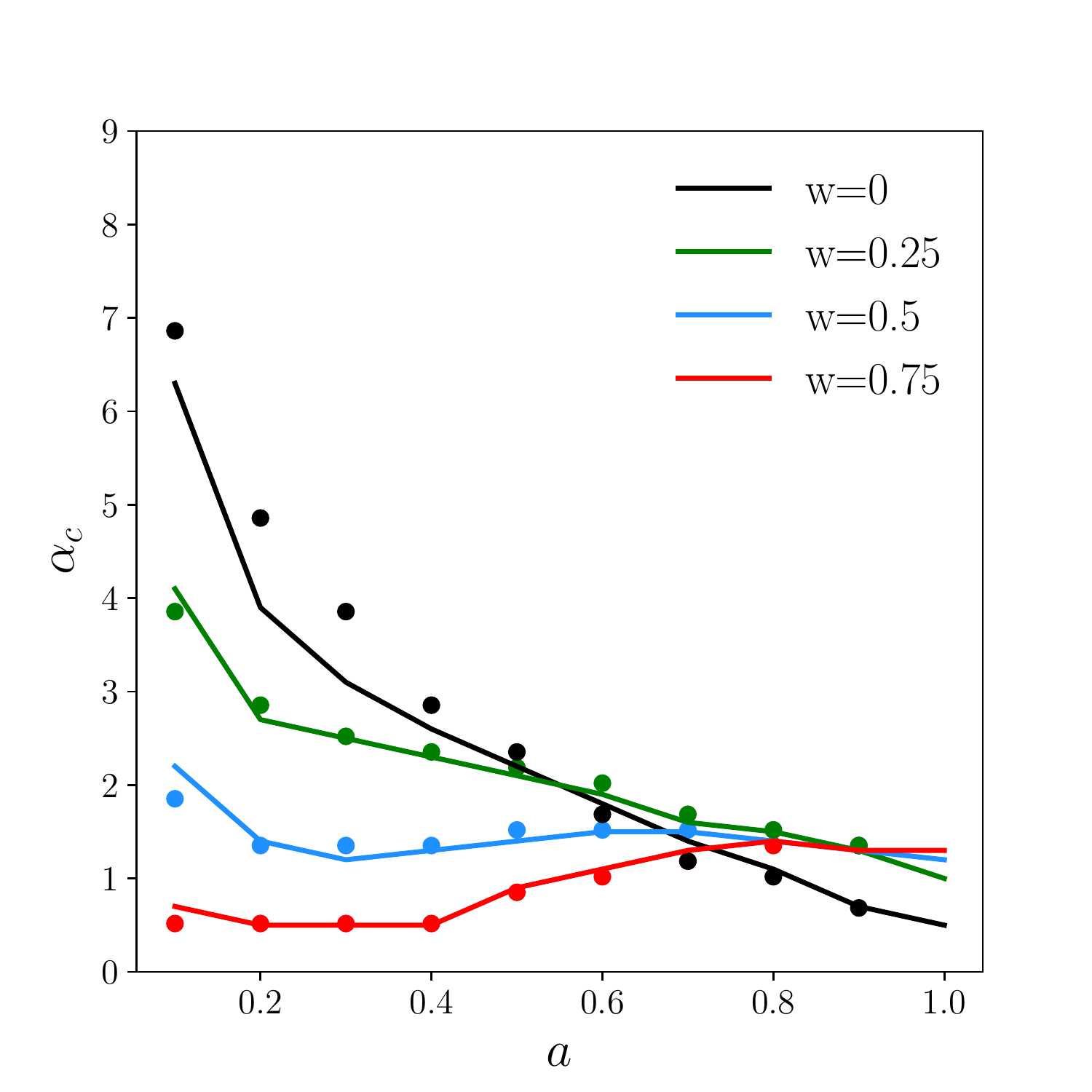}} &
      \hspace*{1cm}
      \resizebox{60mm}{!}{\includegraphics{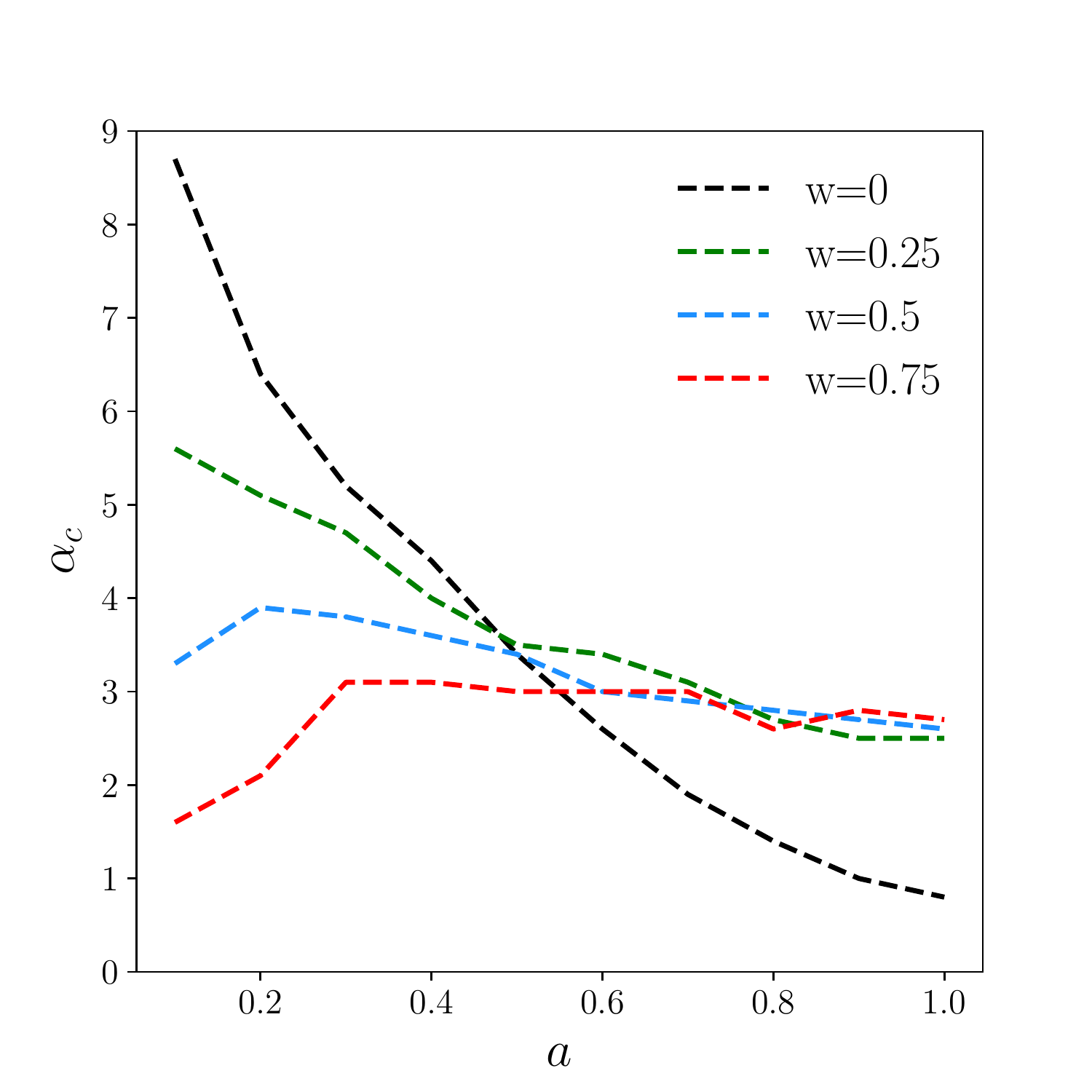}} \\
      &\\
      (a) Fully connected network & (b) Diluted network \\
    \end{tabular}
    \caption{Storage capacity $\alpha_c$ as a function of sparsity $a$ for different values of $w$ for both fully connected (a) and highly (RD) diluted networks (b) as obtained by numerical solution of Eqs.~\eqref{eqn:full_xy}-\eqref{eqn:full_R}. (a) also includes points from simulations. The parameters are $S=5$, $U=0.5$, $\beta=200$.}
    \label{fig:alpha_a_w}
\end{center}
\end{figure}
Fig.~\ref{fig:alpha_a_w} illustrates the same effect of the feedback term, by setting $U=0.5$ and charting the storage capacity as a function of the sparsity $a$ for different values of $w$, for both fully connected (a) and highly diluted networks (b). In both cases, $\alpha_c$ decreases monotonically with increasing $w$, for low $a$, when $U=0.5$ is close to optimal. Increasing $a$, one reaches a region where $U=0.5$ is set too high, and therefore $\alpha_c$ benefits from a non-zero $w$, even though its exact
value is not critical. For very high sparsity parameter (non-sparse coding) all curves except $w=0$ seem to coalesce. The envelope of the different curves represents optimal  threshold setting that takes feedback into account, and as a function of $a$ it shows, both for fully connected and diluted networks the decreasing trend familiar from the analysis of simpler memory networks \cite{TrevesRolls1991}.

\begin{figure}
\begin{center}
    \begin{tabular}{ccc}
      \resizebox{45mm}{!}{\includegraphics{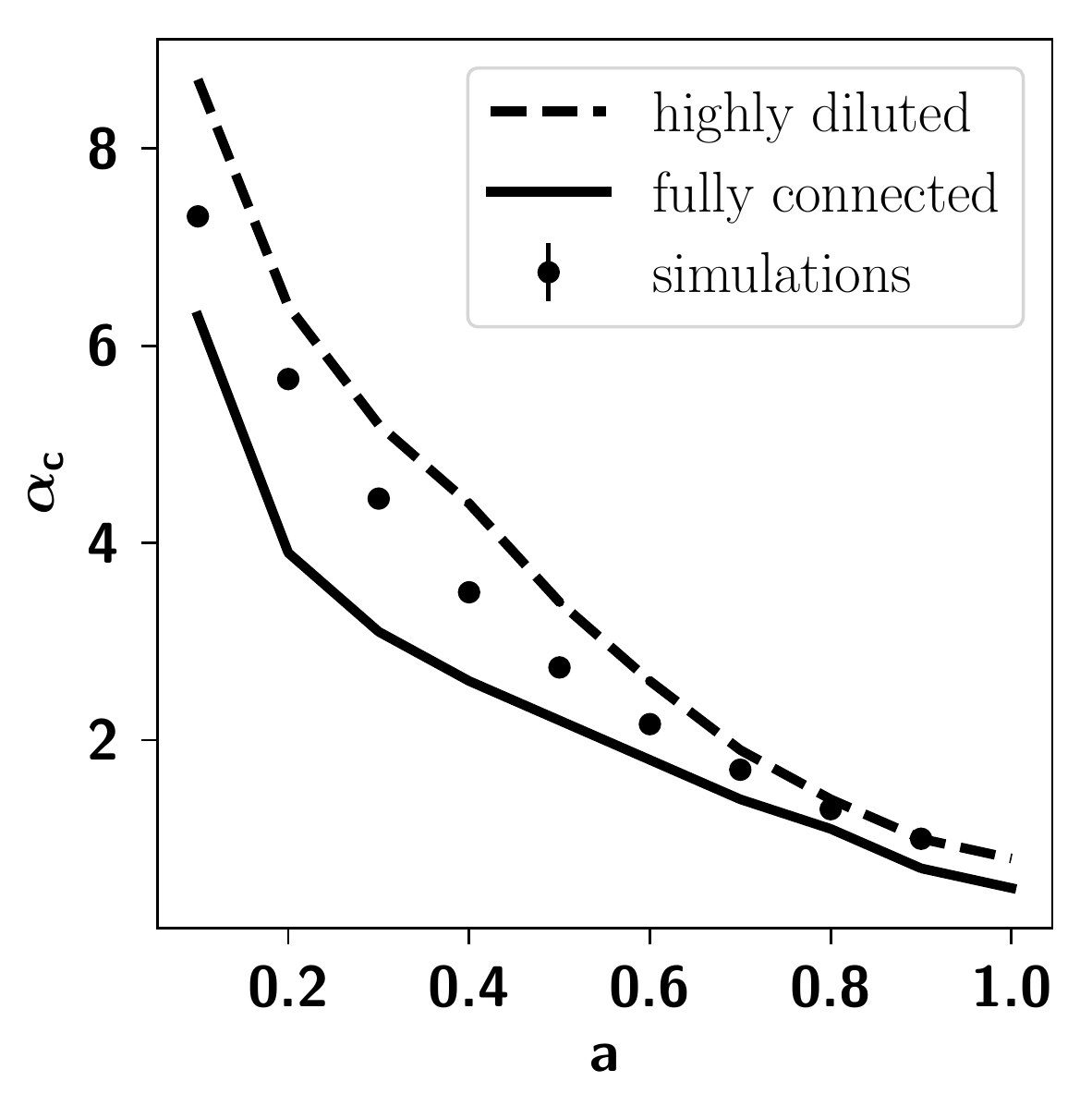}} &
      \resizebox{45mm}{!}{\includegraphics{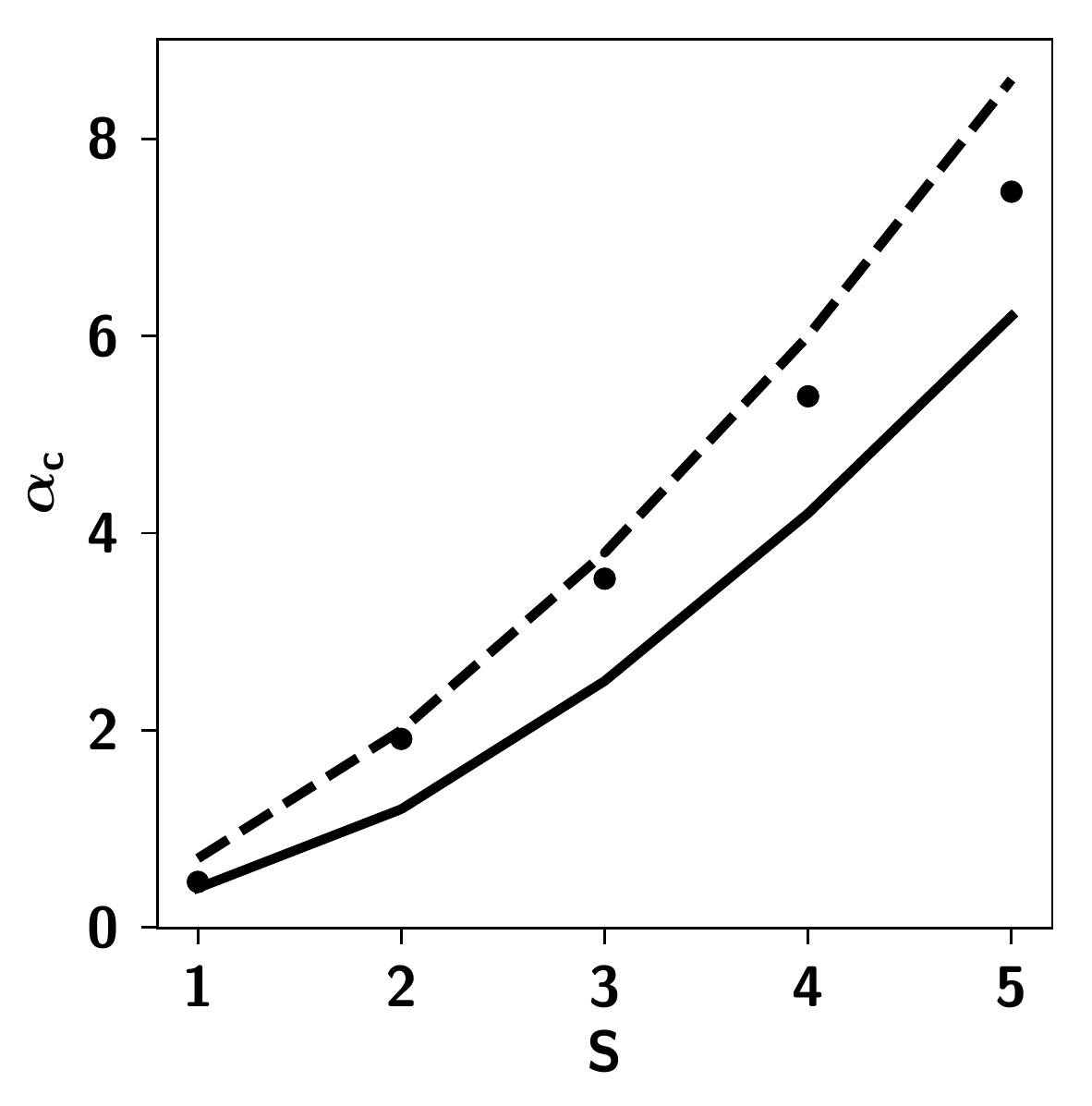}} &
      \resizebox{48mm}{!}{\includegraphics{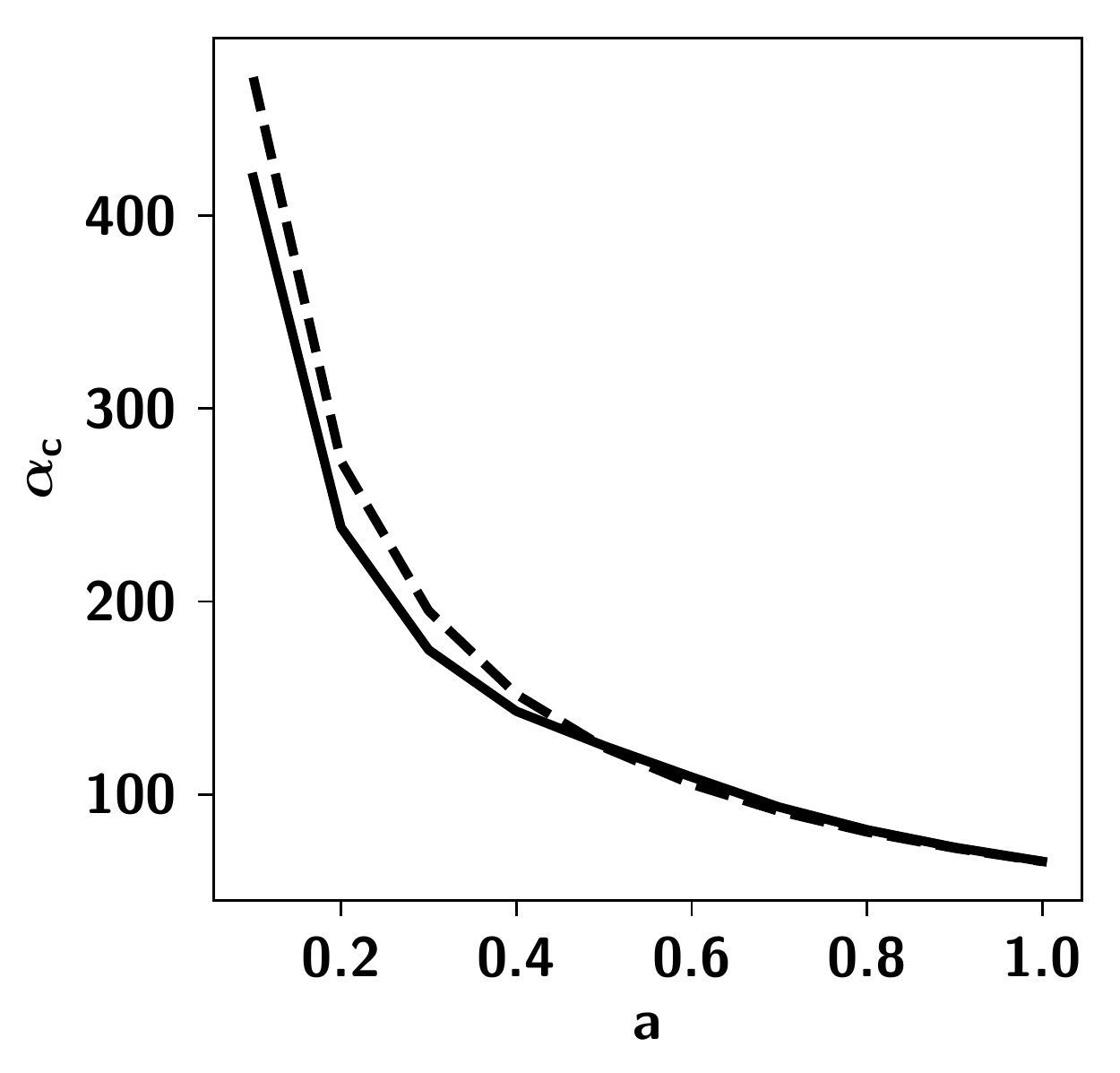}} \\
      (a)  & (b)  & (c) \\
    \end{tabular}
    \caption{(a) Storage capacity $\alpha_c$ as a function of the sparsity $a$. Dots correspond to simulations of a network with $N=2000$, $c_m/N = 0.1$, $S=5$, and $\beta=200$ while curves are obtained by numerical solution of Eqs.~\eqref{eqn:full_xy}-\eqref{eqn:full_R}. (b) Storage capacity as a function of $S$ with the same parameters as in (a) and with $a=0.1$. (c) $S=50$,
illustrating the $\tilde{a}\ll 1$ limit case.}
\label{fig:alpha_a}
\end{center}
\end{figure}
The two connectivity limit cases are illustrated in Fig.~\ref{fig:alpha_a}, which shows, in (a), the dependence of the storage capacity $\alpha$ on the sparsity $a$ in the fully connected and diluted networks with $U=0.5$, $w=0$ and $S=5$. In Fig.~\ref{fig:alpha_a} (b) instead, $S$ is varied and in Fig.~\ref{fig:alpha_a} (c) $S=50$, corresponding to the highly sparse limit $\tilde{a} \ll 1$. While for $S=5$ the two curves are distinct, for the highly sparse network with $S=50$ the two curves coalesce. The
curves are obtained by numerically solving Eqs.~\eqref{eqn:full_xy}-\eqref{eqn:full_R}. Moreover, the storage capacity curve for the fully connected case in (a) matches very well with Fig. 2 of \cite{kropff2005storage}. Diluted curves are always above the fully connected ones in both (a) and (b), as found in \cite{kropff2005storage}.

\subsection{The effect of the different connectivity models}

\begin{figure}[h]
	\centering
    \includegraphics[scale=0.55]{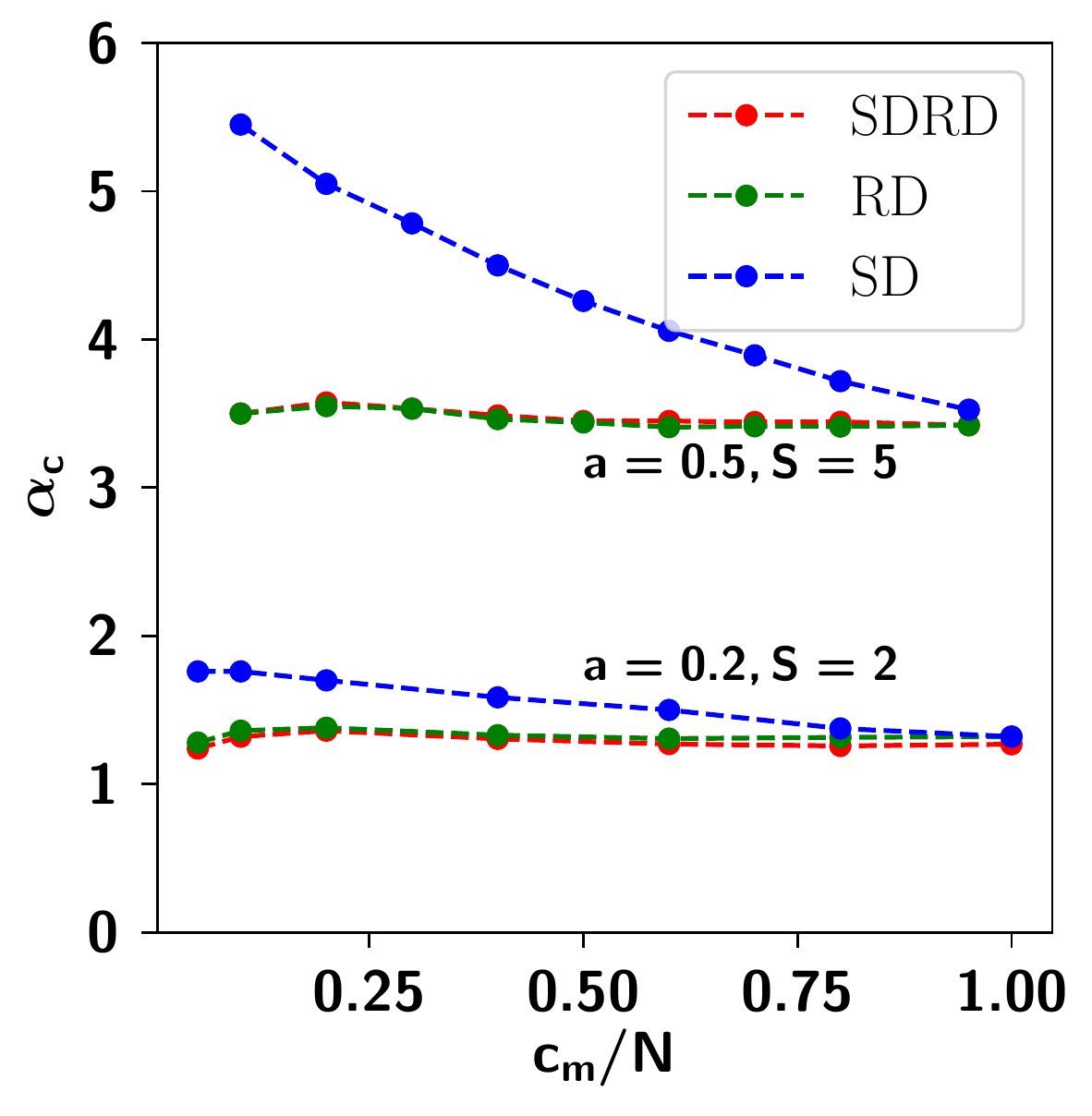}
    \hspace{0.5cm}
    \caption{Storage capacity curves, obtained through simulations, as a function of the mean connectivity per unit $c_m/N$, for three different types of connectivity, namely the random dilution (RD), symmetric dilution (SD) and state-dependent random dilution (SDRD). We find that SD has higher capacity than RD. The capacity for all three models coalesces at the fully connected limit, as the models become equivalent. Simulations carried out for two sets of parameters: ($N=5000$, $S=2$, $a=0.2$) and ($N=2000$, $S=5$, $a=0.5$). $U=0.5$ and $\beta=200$.}\label{fig:SD_RD_SDRD_S2_S5}
\end{figure}

In Fig.~\ref{fig:SD_RD_SDRD_S2_S5} we show simulation results for the storage capacity of all three connectivity models introduced earlier. The RD and SDRD networks seem to have almost identical capacity. All models have the same capacity in the fully connected case, as they should. Note in particular the very limited decrease of $\alpha_c = p/c_m$ with $c_m/N$ increasing up to
almost full connectivity, with all three models. In particular with the RD model, as already shown analytically, the degree of dilution has almost no effect, because already for moderate values of $S$ the network is effectively in the sparse coding regime, $a/S \ll 1$, where $c_m/N$ becomes irrelevant. The apparent decrease in capacity for very low $c_m/N$ values is likely an artefact of $c_m$ being very small.

Our results can be contrasted to the storage capacity with the same connectivity models (RD and SD; SDRD is not relevant) of the Hopfield model. For the Hopfield model, the effects of SD were investigated and it was found that the capacity decreases monotonically from the value $\simeq 0.4$ for highly diluted to the well-known value of $\alpha_c \simeq 0.14$ for the fully connected network \cite{Sompolinsky1986}. In \cite{derrida1987exactly}, instead, the highly diluted limit of RD was studied and a
value of $\alpha_c = 2/\pi \simeq 0.64$ was found. If we plausibly assume that the intermediate RD values interpolate those of the highly diluted $\alpha_c \simeq 0.64$ and fully connected $\alpha_c \simeq 0.14$ limit cases, the Hopfield network seems to have higher capacity for RD than for SD.

However, it is important to note that the overlap with which the network retrieves at $\alpha_c$, $m_c$, is not the same in the two models (RD and SD). In the highly diluted RD model \cite{derrida1987exactly}, the authors find that at zero temperature (which is the only case we consider), $m_c$ undergoes a second order phase transition with control parameter $\alpha$, such that $m_c \simeq \sqrt{3(\alpha_c - \alpha)}$: close to $\alpha_c$, $m_c$ is small, and smaller than the values of $m_c$ for the highly diluted SD model \cite{Sompolinsky1986} that we report in the left $y$-axis of Fig.~\ref{fig:S1_Ui_N2000}: at $c_m/N \simeq 0.0024$, $m_c \simeq 0.64$.
If we require the same precision of retrieval from the RD model, the above equation yielding the $m_c$ gives us a value $\alpha \simeq 0.5$, still higher than the analytic SD value of $0.4$. However, through the simulations shown in Fig.~\ref{fig:S1_Ui_N2000}
we have found that the SD network has a higher capacity ($>0.6$) than the one predicted analytically ($0.4$).

When taking into consideration, for the Hopfield model, the increased capacity of the SD model with respect to what is predicted analytically, as well as the precision of retrieval, we find that the two models behave similarly. We clarify this in the next section
by making the Potts-Hopfield correspondence exact, in a different sense than when considering the multi-modular Hopfield model.

\subsection{The Hopfield model as a special case of the Potts model, for $S=1$}

We can rewrite the Potts Hamiltonian, Eq.~\eqref{eqn:mainham} with $S=1$, $a=0.5$, $U=w=0.0$ such that:
\begin{equation}\label{eqn:binaryham}
	H = -\frac{1}{2} \sum\limits^{N}_{i,j\neq i} J_{ij} \sigma_{i} \sigma_{j} \, ,
\end{equation}
\begin{equation}
	J_{ij} = \dfrac{4}{c_m} \sum\limits^{p}_{\mu=1}\left(\xi_i^\mu - \frac{1}{2}\right)\left(\xi_j^\mu - \frac{1}{2}\right) \, .
\end{equation}
where $\sigma$ and $\xi$ take the values $\lbrace 0,1 \rbrace$. We can rewrite the latter quantities using the spin formulation $\lbrace -1,+1 \rbrace$ using the transformation $2\sigma_i = s_i+1$
\begin{equation}\label{eqn:spinham}
	\tilde{H} = -\frac{1}{8} \sum\limits^{N}_{i,j\neq i} c_{ij} \tilde{J_{ij}} s_{i}s_{j} -\frac{1}{8} \sum\limits^{N}_{i,j\neq i} c_{ij} \tilde{J_{ij}} (s_{i}+s_{j}) -\frac{1}{8} \sum\limits^{N}_{i,j\neq i} c_{ij} \tilde{J_{ij}} \, ,
\end{equation}
\begin{equation}
	\tilde{J_{ij}} = \dfrac{1}{c_m} \sum\limits^{p}_{\mu=1} \eta_i^\mu \eta_j^\mu \, .
\end{equation}
We note now that the first term in Eq.~\eqref{eqn:spinham} is the Hopfield Hamiltonian for storing unbiased patterns, modulo a multiplicative term $1/4$ \cite{amit1985spin}; at zero-temperature, however, an overall rescaling of the energies leaves the statistics of the system unchanged, so that we can consider the first term in Eq.~\eqref{eqn:spinham} as exactly the Hopfield
Hamiltonian. The last term is an additive constant that can be neglected, while the second term can be made to vanish by the addition of a unit dependent threshold term to Eq.~\eqref{eqn:spinham}
\begin{equation}\label{eqn:thresh}
	\tilde{U}_i = \dfrac{1}{8}\sum\limits^{N}_{j(\neq i)} (c_{ij} + c_{ji}) \tilde{J_{ij}}
\end{equation}
or equivalently, to Eq.~\eqref{eqn:binaryham} using the binary formulation
\begin{equation}\label{eqn:binaryhamplusthresh}
	H = -\frac{1}{2} \sum\limits^{N}_{i,j\neq i} J_{ij} \sigma_{i} \sigma_{j}  + \sum\limits^{N}_{i} \left( \frac{1}{4} \sum\limits^{N}_{j (\neq i)} (c_{ij}+ c_{ji}) J_{ij} \right) \sigma_i
\end{equation}

\begin{figure}[t!]
	\centering
    \includegraphics[scale=0.5]{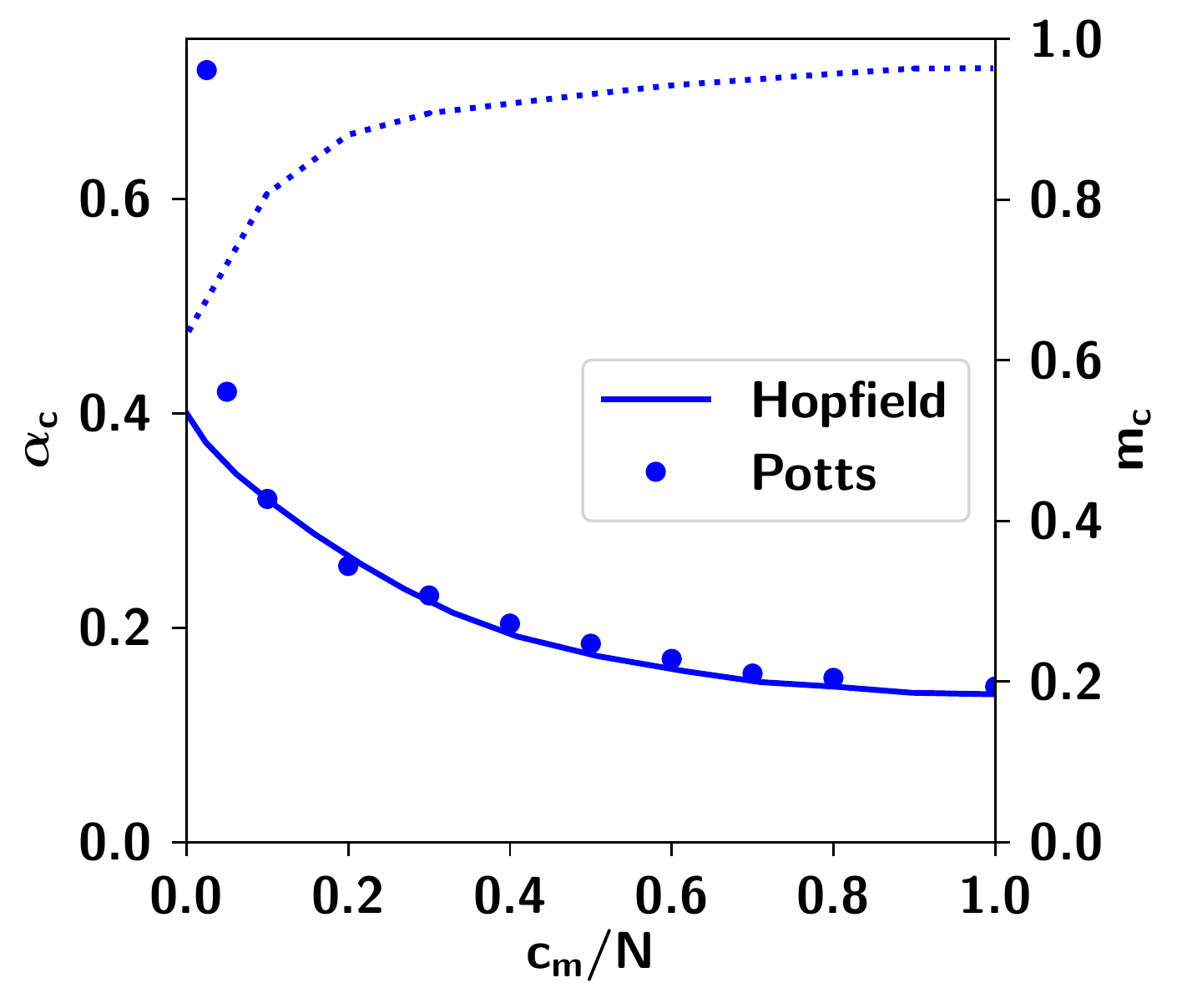}
    \caption{Setting $S=1$, $a=0.5$ and the threshold to be unit dependent ($U=U_i$) the Hamiltonians of the two models become equivalent. Dots correspond to simulations of the Potts network with the latter parameters, while the uninterrupted line corresponds to analytical results obtained by Sompolinsky. The dashed line, to be read with the right $y$-axis, corresponds to the overlap at the critical capacity. For intermediate values of the connectivity, up to $c_m/N=0.1$, our simulation results fit the analytical curve well,
and we find, in particular, the well-known value of $\simeq 0.14$ for the fully connected network. For higher levels of dilution, we find a greater capacity than predicted analytically. Simulations performed with a network of $N=2000$ units.}\label{fig:S1_Ui_N2000}
\end{figure}

Considering $c_{ij}$ to be of the SD type such that  $c_{ij}= c_{ji}$, this is the Hamiltonian considered by Sompolinsky \cite{Sompolinsky1986}. The system with Hamiltonian given by Eq.~\eqref{eqn:binaryhamplusthresh} can be simulated by setting the parameters of the Potts network to $S=1$, $a=0.5$ and $U=U_i$ and the results compared to the analytical results derived in the latter study. We have carried out the simulations to reproduce these results, that we report in Fig.~\ref{fig:S1_Ui_N2000}. Instead, considering $c_{ij}$ to be of the RD type yields the model studied by \cite{derrida1987exactly} at the highly diluted limit.

The unit-dependent threshold that correlates with the learned patterns (and in our case with the diluted connectivity), for the equivalence of the two formulations of the Hamiltonians with spin and binary variables, was first found to be significant when the storage of biased patterns was considered \cite{tsodyks1988enhanced}.

\newpage
\section{Discussion}

In this paper we elaborate on the correspondence between a multi-modular neural network and a coarse grained Potts network, by grounding the Hamiltonian of the Potts model in the multi-modular one. Units are taken to be threshold-linear, in the multi-modular model, and they are fully connected within a module, with Hebbian synaptic weights. Sparse connectivity links units that belong to different modules, via synapses that in the cortex impinge primarily on the apical dendrites, after their axons have travelled through the white matter.

We relate Potts states to the overlap or correlation between the activity state in a module and the local memory patterns, i.e., to weighted combinations of the activity of its threshold-linear units. The long range interactions between the modules then roughly correspond, after suitable assumptions about inhibition, to the tensorial couplings between Potts units in the Potts Hamiltonian.
It becomes apparent how the $w$-term, which was initially introduced by \cite{russo2012cortical} to model positive state-specific feedback on Potts units, arises from the short range interactions of the multi-modular Hamiltonian.

Keeping the $w$-term in the Potts Hamiltonian, we apply the replica method to derive analytically the storage capacity for the fully connected Potts model. A simplified derivation is applied also to the highly diluted connectivity network, while the case with intermediate connectivity is studied by a self-consistent signal-to-noise analysis. The intermediate results smoothly interpolate the limit cases of fully and high diluted networks, but the two limit cases themselves are in fact very similar in capacity, if measured by $
\alpha\equiv p/c_m$, in the sparse coding limit $a\to 0$, a limit which is approached very rapidly in the Potts model, because the relevant parameter is in fact $\tilde{a}\equiv a/S$.
The effect of $w$ term is effectively, in the vicinity of the memory states, reduced to altering the threshold, which leads to the storage capacity being suppressed by this term, if the threshold was originally close to its optimal value. If one assumes that the
threshold is set close to its optimal value \emph{after} taking the feedback term into account, the value $w$ becomes irrelevant for the storage capacity, while it still affects network dynamics \cite{Kang2017}.

\subsection{The storage capacity parameters}
\label{param}

In the end, the storage capacity of the Potts network is primarily a function of a few parameters, $c_m$, $S$ and $a$, that suffice to broadly characterize the model, with minor adjustments due to other factors. How can these parameters be considered to reflect cortically relevant quantities? This a critical issue, if we are to make cortical sense of the distinct thermodynamic phases that can be analysed with the Potts model, and to develop informed conjectures about cortical phase transitions \cite{treves2005frontal}.

The Potts network, if there are $N_{m}$ Potts variables, requires, in the fully connected case, $N_{m} \cdot (N_{m}-1) \cdot S^2/2$ connection variables (since weights are taken to be symmetric we have to divide by 2). In the diluted case, we would have $N_{m} \cdot c_m \cdot S^2$ variables (the factor 2 is no longer relevant, at least for $c_m \to 0$). The multi-modular Hopfield network, as shown in Sect.\ref{sec:HP}, has only $N_{m} \cdot N_{u} \cdot C_A$ long-range synaptic weights.
This diluted connectivity between modules is summarily represented in the Potts network by the tensorial weights. Therefore, the number of Potts weights cannot be larger than the total number of underlying synaptic weights it represents. Then $c_m \cdot S^2$ cannot be larger than $C_A \cdot N_{u}$.

In the simple Braitenberg model of mammalian cortical connectivity \cite{braitenberg1978cortical}, which motivated the multi-modular network model \cite{o1992short}, $N_u\simeq N_m \sim 10^3 - 10^5$, as the total number of pyramidal cells ranges from $\sim 10^6$ in a small mammalian brain to $\sim 10^{10}$ in a large one. In a large, e.g. human cortex, a module may be taken to correspond to roughly $1 \, \textnormal{mm}^2$ of cortical surface, also estimated to include $N_{u} \sim 10^5$ pyramidal cells \cite{braitenberg1978cell}. A module, however, cannot be plausibly considered to be fully connected; the available measures suggest that, even at the shortest distance, the connection probability between pyramidal cells is at most of order 1/10. Therefore we can write, departing from the assumption $C_B= N_u-1$ in the simplest version of Braitenberg's model, that $C_B \simeq 0.1 N_u$. If we were to keep the approximate equivalence $C_A \simeq C_B$, that would imply also $C_A \simeq 0.1 N_u$. Inserting this into the inequality above, $c_m \cdot S^2 < C_A \cdot N_{u}$, yields the constraint $S < N_u \sqrt{0.1/c_m}$.

One can argue, however, for another constraint that limits the value $S$, given the Potts connectivity  $c_m$. The number $S$ of local patterns on $N_{u}$ neurons receiving $C_B$ connections from each other can at most be, given associative storage of patterns with sparsity $a_u$, of order $C_B / a_u$. If we assume that local storage tends to saturate this capacity bound, \emph{and} we take $C_A \simeq C_B$, we have $S \cdot a_u \simeq C_B \simeq C_A $, but again we have, above,   $c_m \cdot S^2 < C_A \cdot N_{u}$, hence
\begin{equation*}
S < N_{u} \cdot a_u / c_m.
\end{equation*}
This more stringent upper bound is compatible with $S$ small and $c_m$ scaling linearly with $N_u$, as well as with $c_m$ small and $S$ scaling with $N_u$, and all intermediate regimes. If we take $S$ and $c_m$ to be both proportional to $\sqrt{N_u}$, and $a_u\sim 0.1$, it would lead to $c_m$ and $S$ to be at most of order $10^1 - 10^2$ over mammalian cortices of different scale, essentially scaling like the fourth root of the total number of pyramidal cells, which appears like a plausible, if rough, modelling assumption.

We could take these range of values, together with the approximate formula (see \cite{kropff2005storage} and Fig.~\ref{fig:alpha_a}b)
\begin{equation}\label{eqn:final_capacity}
p_c \sim 0.15 \frac{c_m S^2}{a\ln (S/a)}
\end{equation}
to yield estimates of the actual capacity the cortex of a given species. The major factor that such estimates do not take into account, however, is the correlation among the memory patterns. All the analyses reported here apply to randomly assigned
memory patterns. The case of correlations will be treated elsewhere (Boboeva, PhD Thesis, SISSA, 2018; and Boboeva
\emph{et al.}, in preparation).

The above considerations may sound rather vague. They neglect, \emph{inter alia}, the large variability in the number of spines, hence probably in synapses, among cortical areas within the same species \cite{Elston}. They capture, however, the quantitative change of perspective afforded by the coarse graining inherent in the Potts model. We can simplify the argument by neglecting sparse coding as well as the exact value of the numerical pre-factor $k$ (which is around 0.15 in Eq.~\eqref{eqn:final_capacity}.
The Potts model uses $N_m c_m S^2$ weights to store up to $k c_m S^2 /\ln S$ memory patterns, each containing of order $N_m \ln S$ bits of information, therefore storing up to $k$ bits per weight. In this respect, and in keeping with the Frolov conjecture \cite{Frolov1997}, it is not different from any other associative memory network based on Hebb-like plasticity, including the
multi-modular model which it effectively represents. In the multi-modular model, however, (in its simplest version) the $2k N_u^2 N_m$ bits available are allocated to memory patterns that are specified in single-neuron detail, and hence contain of order $N_u N_m$ bits of information each. The network can store and retrieve up to a number $p_c$ of them, which has been argued in
\cite{o1992simplest} to be limited by the \emph{memory glass} problem to be of the same order of magnitude as the number $S$ of local attractors, itself limited to be (at most) of order $N_u$ or perhaps, as argued above, $\sqrt{N_u}$.
By glossing over the single-neuron resolution, the Potts model forfeits the locally extensive character of the information contained in each pattern, losing a factor $N_u/\ln S$, but it gains the factor $c_m S^2/(2N_u \ln S)$ in the number of patterns. Whether $S$ scales with $N_u$ or with $\sqrt{N_u}$ or in between, the upshot is more, but less informative, memories. Therefore, by focusing on long range interactions the Potts model misses out in information, but effectively circumvents the memory glass issue, which had
plagued the earlier incarnation of the Braitenberg idea \cite{braitenberg1978cell}, and stores more patterns. How is that possible, if the Potts model is a reduced description of the underlying multi-modular model? The trick is likely in the Hebbian form of the tensor interactions, Eq.~\eqref{eq:tensor_subtract_adivS}, which is \emph{not} a straightforward reduction -- it implies a fine inhibitory regulation that the multi-modular model had not attempted to achieve.
This argument can be expanded and made more precise by considering, again, a more plausible scenario with correlated memories.

Finally, separate studies are needed also to assess the dynamical properties of Potts network, which also reflect the strength of the $w$-term, as we have begun to undertake in an earlier paper elsewhere \cite{Kang2017}. It is such an analysis of the dynamics that may reveal the unique statistical properties of large cortices, as expressed in latching dynamics \cite{treves2005frontal}.

\section*{Acknowledgements}

Work supported by the Human Frontier collaboration with the groups of Naama Friedmann and R{\'e}mi Monasson on analog computations underlying language mechanisms, HFSP RGP0057/2016.

\newpage
\appendix

\section{Calculation of replica symmetric free energy}

The partition function $Z^n$ of $n$ replicas can be written as

\begin{eqnarray}
	 \Big\langle Z^n \Big\rangle &= &\Bigg\langle \textrm{Tr}_{\lbrace \sigma^\gamma \rbrace} \exp \left[ -\beta \sum_\gamma^n H^\gamma \right] \Bigg\rangle \\
		&= &\Bigg\langle \textrm{Tr}_{\lbrace \sigma^\gamma \rbrace} \exp \Bigg\lbrack \frac{\beta}{2Na \left( 1 - \tilde{a} \right)}\sum_{\mu \gamma} \left( \sum_{i}^N v_{\xi_{i}^\mu \sigma_i^\gamma} \right)^2 - \frac{\beta}{2Na \left( 1 - \tilde{a} \right)}\sum_{i}^N \sum_{\mu \gamma} v_{\xi_{i}^\mu \sigma_i^\gamma}^2 \nonumber\\
		&&- \beta \tilde{U} \sum_{i \gamma} \frac{v_{\xi_{i}^\mu \sigma_i^\gamma}}{\delta_{\xi_{i}^\mu \sigma_i^\gamma}- \tilde{a}}  \Bigg\rbrack \Bigg\rangle \, . \nonumber
	\label{eqn:Z1}
\end{eqnarray}
Using the Hubbard-Stratonovich transformation
\begin{equation*}
	\centering
	\exp \left[ \lambda a^2 \right] = \int \frac{dx}{\sqrt{2 \pi}} \exp \left[ - \frac{x^2}{2} + \sqrt{2 \lambda} a x \right] \, ,
\end{equation*}
the first term in Eq.~\eqref{eqn:Z1} can be written as
\begin{equation*}
	\centering
	\exp \left[ \frac{\beta}{2Na \left( 1 - \tilde{a} \right)} \left( \sum_{i}^N v_{\xi_{i}^\mu \sigma_i^\gamma} \right)^2 \right] = \int \frac{d m_\mu^\gamma}{\sqrt{2 \pi}} \exp \left[ - \frac{\left( m_\mu^\gamma \right)^2}{2} + \sqrt{\frac{\beta}{Na \left( 1 - \tilde{a} \right)}} m_\mu^\gamma \sum_{i}^N v_{\xi_{i}^\mu \sigma_i^\gamma} \right] \, .
\end{equation*}

The change of variable $m_\mu^\gamma \rightarrow  m_\mu^\gamma \sqrt{ \beta Na \left( 1 - \tilde{a} \right)} $, and neglecting the sub-leading terms in the $N\rightarrow \infty$ limit, gives us
\begin{eqnarray}
	\Big\langle Z^n \Big\rangle &= &\Bigg\langle \textrm{Tr}_{\lbrace \sigma^\gamma \rbrace} \int \prod_{\mu \gamma} d m_\mu^\gamma \, \cdot \nonumber \\
&&  \cdot \exp \beta N \Bigg\lbrack \frac{a \left( 1 - \tilde{a} \right)}{2} \sum_{\mu \gamma} \left( m_\mu^\gamma \right)^2 + \sum_{\mu \gamma} \frac{m_\mu^\gamma}{N} \sum_{i}^N v_{\xi_{i}^\mu \sigma_i^\gamma} - \frac{1}{2N^2a \left( 1 - \tilde{a} \right)}\sum_{i}^N \sum_{\mu \gamma} v_{\xi_{i}^\mu \sigma_i^\gamma}^2 \nonumber\\
&&-  \frac{1}{N} \tilde{U} \sum_{i \gamma} \frac{v_{\xi_{i}^\mu \sigma_i^\gamma}}{\delta_{\xi_{i}^\mu \sigma_i^\gamma - \tilde{a}}}  \Bigg\rbrack \Bigg\rangle \, .
	\label{eqn:Z2}
\end{eqnarray}

Discriminating the condensed patterns ($\nu$) from non condensed ones ($\mu$) in the limit $p\rightarrow\infty$ and $N\rightarrow\infty$ with the fixed ratio $\alpha=p/N$,

\begin{eqnarray}
 \Big\langle Z^n \Big\rangle &= & \textrm{Tr}_{\lbrace \sigma^\gamma \rbrace} \int \prod_{\mu \gamma} d m_\mu^\gamma \int \prod_{\lambda \gamma} d q_{\gamma \lambda} d r_{\gamma \lambda} \, \cdot \exp \Bigg\lbrace - \frac{\beta N}{2} \sum_{\mu > s} \Bigg\lbrack a \left( 1 - \tilde{a} \right) \sum_\gamma \left( m_\mu^\gamma \right)^2\nonumber \\
 & &  -  a \left( 1 - \tilde{a} \right) \beta \tilde{a} \sum_{\gamma \lambda} m_\mu^\gamma m_\mu^\lambda q_{\gamma \lambda} \Bigg\rbrack - \frac{\alpha \beta \tilde{a} N}{2} \sum_{\gamma \gamma} q_{\gamma \gamma}-\beta N a \tilde{U} \sum_{\gamma \gamma} q_{\gamma \gamma} \nonumber \\
  & &  - \frac{N \alpha \beta^2}{2} \sum_{\gamma \lambda} r_{\gamma \lambda} \left( \tilde{a}^2 q_{\gamma \lambda} - \frac{1}{N S \left( 1 - \tilde{a}\right)} \sum_{i k} P_k v_{k \sigma_i^\gamma} v_{k \sigma_i^\lambda} \right) \Bigg\rbrace \cdot \Bigg\langle \exp \beta N \Bigg\lbrack \frac{a ( 1 - \tilde{a} )}{2}  \nonumber\\
  &&  \sum_{\nu \gamma}^{\nu \leq s} \Big( m_\nu^\gamma \Big)^2 + \sum_{\nu \gamma}^{\nu \leq s} \frac{m_\nu^\gamma}{N} \sum_{i}^N v_{\xi_{i}^\nu \sigma_i^\gamma} - \frac{1}{2 N^2 a ( 1 - \tilde{a} )}\sum_{i}^N \sum_{\nu \gamma}^{\nu \leq s} v_{\xi_{i}^\nu \sigma_i^\gamma}^2 \Bigg\rbrack \Bigg\rangle
\label{eqn:Z6}
\end{eqnarray}
where we introduced $q_{\gamma \lambda}$, the overlap between different replicas, analogous to the Edwards-Anderson order parameter \cite{edwards1975theory},
\begin{equation}
	\centering
	q_{\gamma \lambda} = \frac{1}{N a \tilde{a} \left( 1 - \tilde{a}\right)} \sum_{i k} P_k v_{k \sigma_i^\gamma} v_{k \sigma_i^\lambda} \, .
	\label{eqn:qEA}
\end{equation}
The saddle point equations are
\begin{equation}
	\centering
	\frac{\partial}{\partial m_\nu^\gamma} = 0 \longrightarrow m_\nu^\gamma = \Bigg\langle \frac{1}{N a \left( 1 - \tilde{a}\right)} \sum_{i} \left\langle v_{\xi_i^\nu \sigma_i^\gamma} \right\rangle \Bigg\rangle \, ,
	\label{eqn:m_nu}
\end{equation}
\begin{equation}
	\centering
	\frac{\partial}{\partial r_{\gamma \lambda}} = 0 \longrightarrow q_{\gamma \lambda} = \frac{1}{N a \tilde{a} \left( 1 - \tilde{a}\right)} \sum_i^N \Bigg\langle \sum_k P_k \left\langle v_{k \sigma_i^\gamma} v_{k \sigma_i^\lambda} \right\rangle \Bigg\rangle \, ,
	\label{eqn:q_gammalambda}
\end{equation}
\begin{equation}
	\centering
	\frac{\partial}{\partial q_{\gamma \lambda}} = 0 \longrightarrow r_{\gamma \lambda} = \frac{S \left( 1 - \tilde{a} \right) }{\alpha} \sum_\mu \Bigg\langle  m_\mu^\gamma  m_\nu^\lambda  \Bigg\rangle - \left[ \frac{2 S}{\alpha} \tilde{U}  + 1 \right] \frac{\delta_{\gamma \lambda}}{\beta \tilde{a}} \, .
	\label{eqn:r_gammalambda}
\end{equation}

After performing the multidimensional Gaussian integrals over fluctuating (non condensed) patterns we have
\begin{eqnarray}
	\Big\langle Z^n \Big\rangle& = & \int \prod_{\nu \gamma}^{\nu \in \left[ 1, \dots, s \right]} d m_\nu^\gamma \int \prod_{\lambda \gamma} d q_{\gamma \lambda} d r_{\gamma \lambda} \, \cdot \nonumber \\
	 &&  \cdot \exp N \Bigg\lbrace - \beta \frac{a \left( 1 - \tilde{a} \right) }{2} \sum_{\nu \gamma} \left( m_\nu^\gamma \right)^2 -  \frac{\alpha}{2} \textrm{Tr} \ln \left[ a \left( 1 - \tilde{a} \right) \left(1 - \beta \tilde{a} \textbf{q} \right) \right] - \\
	 & &\frac{\alpha \beta^2 \tilde{a}^2}{2} \sum_{\gamma \lambda} r_{\gamma \lambda} q_{\gamma \lambda} - \beta \tilde{a} \left[ \frac{\alpha}{2} + S \tilde{U} \right] \sum_{\gamma \gamma} q_{\gamma \gamma} + \Bigg\langle \ln \textrm{Tr}_{\lbrace \sigma^\gamma \rbrace} \exp \left[ \beta \mathcal{H}_\sigma^\xi \right] \Bigg\rangle_{\xi^v} \Bigg\rbrace \,  \nonumber,
	\label{eqn:Z7}
\end{eqnarray}
where
\begin{equation}
	\centering
	\mathcal{H}_\sigma^\xi = \sum_{\nu \gamma} m_\nu^\gamma v_{\xi^\nu \sigma^\gamma} + \frac{\alpha \beta}{2 S\left( 1 - \tilde{a} \right)} \sum_{\gamma \lambda} r_{\gamma \lambda} \sum_k P_k v_{k \sigma^\gamma} v_{k \sigma^\lambda} \, .
\end{equation}

We can now compute the free energy Eq.~\eqref{eqn:ftrick}
\begin{eqnarray}
	f &= & \nonumber \lim_{n \rightarrow 0} f_n = \lim_{n \rightarrow 0}  \Bigg\lbrace \frac{a \left( 1 - \tilde{a} \right) }{2 n} \sum_{\nu \gamma} \left( m_\nu^\gamma \right)^2 + \nonumber \\
	& +&  \frac{\alpha}{2 n \beta} \textrm{Tr} \ln \left[ a \left( 1 - \tilde{a} \right) \left(1 - \beta \tilde{a} \textbf{q} \right) \right] + \frac{\alpha \beta \tilde{a}^2}{2n} \sum_{\gamma \lambda} r_{\gamma \lambda} q_{\gamma \lambda} \nonumber \\
	& +& \frac{\tilde{a}}{n} \left[ \frac{\alpha}{2} + S \tilde{U} \right] \sum_{\gamma \gamma} q_{\gamma \gamma} - \frac{1}{n\beta} \Bigg\langle \ln \textrm{Tr}_{\lbrace \sigma^\gamma \rbrace} \exp \left[ \beta \mathcal{H}_\xi \right] \Bigg\rangle_{\xi^v} \Bigg\rbrace \, .
	\label{eqn:ffinal}
\end{eqnarray}
Imposing the replica symmetry condition \cite{sherrington1975solvable},
\begin{eqnarray*}
 		 m_{\gamma}^\nu &=& m\\
 		  q_{\gamma \lambda}&=&
 		  \left\{
 		  \begin{array}{lll}
 		  q & \textrm{for} \,& \gamma \neq \lambda \\
       \tilde{q} & \textrm{for} \,& \gamma = \lambda\\
       \end{array}
       \right.\\
        r_{\gamma \lambda} &=&
        \left\{
        \begin{array}{lll}
 		r & \textrm{for} \,& \gamma \neq \lambda \\
       \tilde{r} & \textrm{for} \,& \gamma = \lambda
       \end{array}
       \right.
\end{eqnarray*}
we finally obtain the replica symmetric free energy Eq.~\eqref{eqn:freplicasymmetric}.
\newpage
\section{Self consistent signal to noise analysis}
Since the l.h.s. of Eq.~\eqref{eqn:ansatz_noise_intermediate} includes $p-1 \gg 1$ terms,
the ansatz is still valid also when singling out one of these many contributions, so that we can equivalently write it as
\begin{equation}
	\sum_{\nu>1} v_{\xi_i^{\nu},k} m_i^{\nu}
		= v_{\xi_i^{\mu},k} m_i^{\mu} + \sum_{\nu\neq 1,\mu} v_{\xi_i^{\nu},k} m_i^{\nu}
		= v_{\xi_i^{\mu},k} m_i^{\mu} + \gamma_i^k \langle \sigma_i^k \rangle + \sum_n v_{n,k} \, \rho_i^n z_i^n \ ,
\end{equation}
where $\gamma_i^k$ and $\rho_i^n$ are independent of $\mu$.
The contribution from the non-condensed pattern $\mu\neq 1$ is assumed to be small, so that we can expand $G_i^k$ to first order in $v_{\xi_i^{\mu},k} m_i^{\mu}$:
\begin{eqnarray}\label{eqn:Taylor_sigma}
	\sigma_j^l &= & G^l\bigg[\Big\{ v_{\xi_j^{1},k} m_j^{1} + \sum_n v_{n,k} \rho_j^n z_j^n - U(1-\delta_{k,0})\Big\}_{k=0}^S\bigg] \nonumber \\
	&&\qquad + \sum_n v_{\xi_j^{\mu},n} \, m_j^{\mu} \, \frac{\partial G^l}{\partial y^n} \bigg[ \Big\{ v_{\xi_j^{1},k} m_i^{1} + \sum_n v_{n,k} \rho_j^n z_j^n - U(1-\delta_{k,0}) \Big\} \bigg] \ .
\end{eqnarray}
Reinserting the expansion into the r.h.s of Eq.~\eqref{eqn:local_overlap}
we recognize a relation of the form
\begin{equation}\label{eqn:m_function_G_G'}
m_i^{\mu} = L_i^{\mu} + \sum_j K_{ij}^{\mu} m_j^{\mu} \,
\end{equation}
where
\begin{eqnarray*}\label{eqn:K_L}
 K_{ij}^{\mu} &\equiv & \frac{1}{c_m a(1-\tilde{a})} \sum_{l,n} c_{ij} v_{\xi_j^{\mu},l} v_{\xi_j^{\mu},n}  \frac{\partial G_j^{l}}{\partial y^n} \ ,  \\
 L_i^{\mu}& \equiv &\frac{1}{c_m a(1-\tilde{a})} \sum_j \sum_l c_{ij} v_{\xi_j^{\mu},l} G_j^l  \ .
\end{eqnarray*}
The overlap $m_i^\mu$ can be found by iterating Eq.~\eqref{eqn:m_function_G_G'},
\begin{equation}
m_i^{\mu} = L_i^{\mu} + \sum_{j_1} L_{j_1}^{\mu} \bigg\lbrace K_{ij_1}^{\mu} + \sum_{j_2} K_{ij_2}^{\mu} K_{j_2 j_1}^{\mu} + \sum_{j_2} \sum_{j_3} K_{ij_2}^{\mu} K_{j_2 j_3}^{\mu} K_{j_3 j_1}^{\mu} + ... \bigg\rbrace \, .
\end{equation}
Therefore, the noise term can be written explicitly as
\begin{eqnarray*}
	&&\sum_{\mu>1} v_{\xi_i^{\mu}, k} m_i^{\mu} = \sum_n v_{n,k} \sum_{\mu>1} \Bigg\lbrace \sum_j \sum_l \frac{1}{c_ma(1-\tilde{a})} \, c_{ij} \delta_{\xi_i^{\mu},n} v_{\xi^{\mu}_j,l} G_j^l + \\
& +&  \sum_{j_1} \sum_{j} \sum_l \frac{1}{c_ma(1-\tilde{a})} \, c_{j_1 j} \delta_{\xi_i^{\mu},n} v_{\xi_j^{\mu},l} G_j^l \Bigg( \sum_{l_1, n_1} \frac{1}{c_ma(1-\tilde{a})} \, c_{i j_1} v_{\xi_{j_1}^{\mu},l_1} v_{\xi_{j_1}^{\mu},n_1} \frac{\partial G_{j_1}^{l_1}}{\partial y^{n_1}} + ... \Bigg) \Bigg\rbrace \, .
\end{eqnarray*}
In order to obtain the expression for $\gamma_i^k$, in Eq.~\eqref{eqn:ansatz_noise_intermediate}
we consider only the terms with $j=i$ and $l=k$,
and take the average over the connectivity and the patterns:
\begin{eqnarray}
	\gamma_i^k &= & \frac{\alpha}{S} \lambda \bigg\langle \frac{1}{S} \frac{1}{N}\sum_{j_1} \sum_{l_1} \frac{\partial G_{j_1}^{l_1}}{\partial y^{l_1}}  + ... \bigg\rangle \\ \nonumber
&=&\frac{\alpha}{S} \lambda \Big\lbrace \Omega/S +( \Omega/S)^2 + ... \Big\rbrace \\ \nonumber
&=&\frac{\alpha}{S} \lambda\frac{\Omega/S}{1-\Omega/S}
\end{eqnarray}
where we use the fact that $c_{ii}=0$, $\alpha = p/c_m$, $\langle\cdot\rangle$ indicates the average over all patterns and where we have defined
\begin{equation}
\Omega = \bigg\langle\frac{1}{N}\sum_{j_1} \sum_{l_1}\frac{\partial G_{j_1}^{l_1}}{\partial y^{l_1}}  \bigg\rangle\, .
\end{equation}
By virtue of the statistical independence of units, the average over the non-condensed patterns for the $i \neq j$ terms vanishes. From the variance of the noise term one reads
\begin{equation}
(\rho_i^n)^2  =  \frac{\alpha P_n}{S(1-\tilde{a})} q \Big\lbrace 1+ 2\lambda \Psi + \lambda \Psi^2  \Big\rbrace \, ,
\end{equation}
where
\begin{equation}
q = \bigg\langle \frac{1}{Na} \sum_{j,l} (G_j^l)^2 \bigg\rangle
\end{equation}
and
\begin{equation}
\Psi = \frac{\Omega/S}{1-\Omega/S} \ .
\end{equation}

The mean field received by a unit is then
\begin{equation}
 \mathcal{H}_k^{\xi} = v_{\xi,k} m + \frac{\alpha}{S} \lambda \Psi (1- \delta_{k,0}) + \sum_n v_{n,k} z^n \sqrt{\frac{\alpha P_n}{S(1-\tilde{a})} q \Big\lbrace 1+ 2\lambda \Psi + \lambda \Psi^2  \Big\rbrace} - \tilde{U}(1-\delta_{k,0}) \, .
\end{equation}


\end{document}